\documentclass[preprint,10pt]{elsarticle}

\usepackage{graphicx}
\usepackage{amsmath, amsthm, mathtools, amssymb, color, float, eqnarray, booktabs, makecell, soul, mdframed, multirow}
\usepackage[hyphens]{url}
\usepackage[justification=centering]{caption}
\usepackage{todonotes}
\usepackage{comment}
\usepackage{bm}
\usepackage{ulem}
\usepackage[colorlinks=true, linkcolor=blue, citecolor=blue, urlcolor=blue]{hyperref}

\usepackage{cleveref}
\usepackage{subcaption} 
\usepackage[utf8]{inputenc}  
\usepackage[T1]{fontenc}     
\usepackage{enumitem}
\usepackage[most]{tcolorbox}

\usepackage[utf8]{inputenc}
\usepackage{soul}

\journal{arXiv.org}

\begin{document}
\begin{frontmatter}

\title{CMINNs: Compartment Model Informed Neural Networks - Unlocking Drug Dynamics}

\author[1]{Nazanin Ahmadi Daryakenari}
\author[3]{Shupeng Wang}
\author[3]{George Karniadakis\corref{corresponding}}
\address[1]{Center for Biomedical Engineering, Brown University, Providence, RI, USA}
\address[3]{Division of Applied Mathematics, Brown University, Providence, RI, USA}



\cortext[corresponding]{Corresponding author: george\_karniadakis@brown.edu}

\address{}

\begin{abstract}

In the field of pharmacokinetics and pharmacodynamics (PKPD) modeling, which plays a pivotal role in the drug development process, traditional models frequently encounter difficulties in fully encapsulating the complexities of drug absorption, distribution, and their impact on targets. Although multi-compartment models are frequently utilized to elucidate intricate drug dynamics, they can also be overly complex. To generalize modeling while maintaining simplicity, we propose an innovative approach that enhances PK and integrated PK-PD modeling by incorporating fractional calculus or time-varying parameter(s), combined with constant or piecewise constant parameters.
These approaches effectively model anomalous diffusion, thereby capturing drug trapping and escape rates in heterogeneous tissues, which is a prevalent phenomenon in drug dynamics. Furthermore, this method provides insight into the dynamics of drug in cancer in multi-dose administrations. Our methodology employs a Physics-Informed Neural Network (PINN) and fractional Physics-Informed Neural Networks (fPINNs), integrating ordinary differential equations (ODEs) with integer/fractional derivative order from compartmental modeling with neural networks. This integration optimizes parameter estimation for variables that are time-variant, constant, piecewise constant, or related to the fractional derivative order.
The results demonstrate that this methodology offers a robust framework that not only markedly enhances the model's depiction of drug absorption rates and distributed delayed responses but also unlocks different drug-effect dynamics, providing new insights into absorption rates, anomalous diffusion, drug resistance, peristance and pharmacokinetic tolerance, all within a system of just two (fractional) ODEs with explainable results.
\end{abstract}


\end{frontmatter}


\section{Introduction}
Understanding the relationship between dose, concentration, and effect is a cornerstone of clinical pharmacology. This relationship is interpreted through pharmacokinetic (PK) and pharmacodynamic (PD) models. The development of drugs relies on an iterative process of pharmacometric model-informed learning to optimize dosing for both safety and efficacy, spanning from preclinical studies to Phase III clinical trials \cite{Standing2017}. Beyond the realm of drug development, PKPD principles are crucial for medication prescription and administration. These principles are particularly beneficial for determining off-label dosages, personalizing treatments based on biomarkers, and adjusting doses to manage efficacy and toxicity concerns. In clinical research, PKPD models are essential for ensuring optimal dosing and accurate power calculations, thereby enhancing the design and outcomes of studies \cite{Kalaria2022}.
Pharmacometric modeling and simulation offer profound insights into the dose–concentration–effect relationship. A population approach in pharmacometrics, employing mixed effects or multilevel modeling, is used to estimate parameter values while accounting for individual variability \cite{DiStefano1984}. This approach is applicable to both extensive and limited datasets, with optimal design techniques enhancing data collection and precision \cite{Dokoumetzidis2010}. Population pharmacometric models generate typical population parameters and variances, which can facilitate new hypotheses and applications in personalized medicine \cite{Popovic2010}. Utilizing PKPD model parameters as endpoints in clinical trials can improve study power and efficiency. For example, viral kinetic models in hepatitis C have demonstrated increased power in detecting differences in drug effects \cite{Verotta2010}. Investigator-led studies benefit from PKPD in designing dosing guidelines, as shown by the use of insulin-like growth factor (IGF-1) in children with Crohn’s disease \cite{Rao2013}.\\
\textbf{Compartment Models:} Given the complexity of the human body, drug kinetics are often simplified by representing the body as one or more tanks or compartments, which are reversibly interconnected. A compartment is considered a group of tissues with similar blood flow, but it is not an actual anatomical or physiological region. Within each compartment, the drug is assumed to be uniformly distributed \cite{pharmacokinetic2016}.\\
\textbf{Non-Compartment Models:} Non-compartmental analysis, also known as the model-independent approach, does not rely on the assumption of a specific compartmental model. Instead, it assumes that the individual or organism can be represented as a single homogeneous compartment. This approach presumes that the blood-plasma concentration of the drug accurately reflects its concentration in other tissues and that drug elimination is directly proportional to its concentration in the individual or organism \cite{DiStefano1984}.\\

\textbf{Fractional Calculus in Pharmacokinetics and Pharmacodynamics Modeling:} Diffusion plays a crucial role in various transport processes within living organisms and significantly affects drug distribution in the body. Key processes such as membrane permeation, solid dissolution, and dispersion in cellular matrices are primarily driven by diffusion. Traditionally, diffusion is explained by Fick's first law. However, recent experimental evidence has indicated that this conventional understanding does not always hold true. Instances of diffusion that deviate from this norm have been observed, manifesting as either accelerated (super-diffusion) or decelerated (sub-diffusion) relative to the standard case \cite{West1994, Ionescu2017}. In the context of pharmacokinetics, results in nonexponential tissue trapping and exponential washout curves characterized by a specific time scale, typically referred to as a half-life \cite{Weiss1999TheAP}. These atypical diffusion patterns are termed anomalous, as they deviate from the usual diffusion dynamics \cite{West1994}. 
The concept of fractional kinetics was first introduced into pharmaceutical literature by \cite{Dokoumetzidis2011}, with amiodarone being the initial drug demonstrating power-law kinetics \cite{Tucker1984}. Following this, various other applications of fractional pharmacokinetics have emerged in the literature \cite{sopasakis2018fractional, hennion2013avoid}. There has been growing interest in using fractional calculus in pharmacokinetics, where the Mittag-Leffler function offers an alternative to monoexponential models for representing single-compartment kinetics after a bolus dose \cite{Dokoumetzidis2009}. This approach was further expanded to commensurate two-compartment models \cite{Popovic2010}. Additionally, \cite{Verotta2010} highlighted that sums of single- and two-parameter Mittag-Leffler functions can represent the response functions for various, both commensurate and non-commensurate, compartmental systems. These sums, combined with arbitrary inputs, can model a wide range of fractional pharmacokinetics scenarios. The work by \cite{Dokoumetzidis2011} shifted focus towards pharmacodynamics, introducing a novel method for pharmacodynamics modeling using fractional integrals and differential equations. Despite the increasing number of applications of fractional order integrals and differential equations in fields such as physics, signal processing, engineering, and bioengineering, they have received limited attention in pharmacokinetics and pharmacodynamics literature. One reason is the computational challenge: even though analytical solutions to fractional differential equations are available in special cases, the simplest PKPD models constructed using fractional calculus often do not permit analytical solutions \cite{Dokoumetzidis2011}. Unlike systems defined by a single ordinary differential equation (ODE), fractional multi-compartmental models cannot be simply created by changing the order of the ordinary derivatives in the ODEs to fractional orders. This simplistic modification can result in inconsistent systems that violate mass balance. The authors of \cite{Dokoumetzidis2010} present a rationale for the fractionalization of ODEs, ensuring the creation of consistent systems that accommodate processes of different fractional orders within the same model. Recently, new fractional PKPD models have been proposed and compared with traditional integer-order models \cite{Borkor2023InvestigationOF, Miskovic2023, COPOT202127, ZAITRI2023107679, Alinei-Poiana2023, GHITA202161,verotta2010fractional}, highlighting the significance of this approach in situations where the diffusion rate does not exhibit exponential decay or where the number of parameters/compartments in the model is reduced.\\

\textbf{Drug Resistance, Tolerance, and Persistence:} 
Overcoming chemotherapy resistance, tolerance, and persistence is critical for effective cancer treatment, as these mechanisms enable tumor cells to survive and thrive even under intensive therapies. Resistance, typically driven by genetic mutations, allows cancer cells to grow at high drug concentrations, while tolerance and persistence refer to the survival of cells under transient or prolonged drug exposure without genetic changes. The ambiguity and overlap in the definition of these terms complicates the interpretation of patient responses to treatment and underscores the need for precise diagnostic and therapeutic strategies \cite{Brauner2016, Balaban2019, Kulkarni2022}. Recent advances, such as the development of computational models such as Re-sensitizing Drug Prediction (RSDP) and the exploration of adaptive therapeutic strategies, highlight the importance of addressing both genetic and non-genetic factors in drug resistance \cite{WANG2023107230, MASUD2023107035, PEREZALIACAR2024108866, CHEN2024}. The emergence of drug-tolerant persister (DTP) cells, which are slow-growing and metabolically distinct, further complicates treatment outcomes by acting as a reservoir for drug-resistant mutations \cite{He2024, Liang2023, Tyner2022}. These DTP cells, originally identified in antibiotic-resistant bacterial biofilms, lack underlying genomic alterations yet contribute significantly to treatment resistance in cancer \cite{Pu2023}. Understanding the parallels between bacterial and cancer persisters, along with the role of tumor heterogeneity and microenvironmental adaptations, is critical to developing more effective, personalized cancer therapies that can overcome resistance and improve patient outcomes \cite{Housman2014, Lin2024, Thomas2020, Ramos2021, Kulkarni2022, Bell2020, Shlyakhtina2021, Sharma2010, CAI2023100962}.\\

\textbf{Physics-Informed Neural Networks: }
Physics-Informed Neural Networks (PINNs) \cite{Raissi2019} have revolutionized parameter estimation, especially for systems with complex nonlinear dynamics, numerous unknown parameters, and limited experimental data availability \cite{raissi2024physics}. The advent of fractional PINNs (fPINNs) \cite{fPINNs} has broadened these techniques to encompass fractional differential equations, thereby broadening the scope of modeling possibilities. The AI-Aristotle framework \cite{Ahmadi2024} introduced PINNs into pharmacokinetic models, creating a benchmark for addressing complex inverse problems through these advanced models, and a recent paper \cite{Podina2024UPINNs} used PINNs for learning chemotherapy drug action. In the present work, we extend signifucantly this groundwork, by developing new PINN and fPINN models and applying them to diverse cases. This is achieved by modifying the compartment model to fractional or time-varying parameter form, or by simplifying the system to a system of two ODEs. This allows for enhanced precision and the generation of new insights within PK and PD modelling.\\

\textbf{Proposed Methodology: }In pharmacokinetic modeling, we typically favor simpler models, largely due to the fact that we usually have access only to data from the central compartment, namely plasma. Incorporating additional compartments complicates the model by increasing the number of parameters that need to be inferred from the available data. This is particularly problematic in cases where the resulting system is not structurally identifiable. Herein, we introduce two distinct approaches within the physics-informed neural network framework. These methods either maintain the simplicity of the model or reduce the number of compartments while still achieving a good fit to the data. Additionally, we offer new insights by presenting the diffusion rate, rate of cell death, and drug efficacy index as parameters that change over time. This study presents an innovative concept utilizing neural networks for compartmental pharmacokinetic (PK) modeling, effectively identifying all necessary parameters even with limited and sparse datasets.
We compare two primary approaches across three distinct pharmacokinetic compartment models: a two-compartment model with a single diffusion rate, a standard two-compartment model, and a three-compartment model. Each of these models is applied to the pharmacokinetic analysis of separate drug concentration datasets. Additionally, we introduce a fourth model that integrates pharmacokinetics and pharmacodynamics, focusing on the pharmacodynamic aspects to assess cancer drug resistance and drug tolerance in a multi-dose administration regimen, while simplifying the model to a system of two ODEs.

\begin{figure}[!ht]
    \centering
    \includegraphics[width=0.8\linewidth]{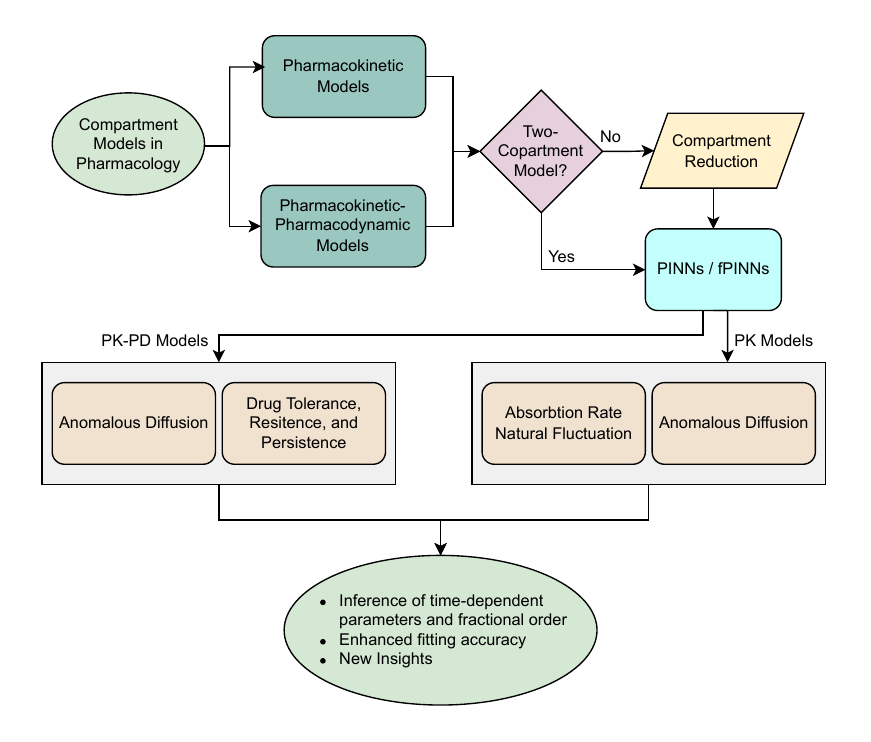}
    \caption{\textbf{CMINNs 
    workflow.} We consider two types of compartment models in pharmacology: pharmacokinetic models and pharmacokinetic-pharmacodynamic models. When the model is represented by a system of two ODEs, we apply the PINNs/fPINNs framework to introduce additional flexibility in capturing complex dynamics. This enhances fitting accuracy, particularly for models with reduced parameters or non-exponential decay dynamics, where traditional methods fail to achieve satisfactory accuracy and insights. For multi-compartment models, the workflow proceeds to a compartment reduction step, reformulating the model to optimize accuracy and provide deeper insights into drug dynamics. This allows for comparative analysis of different drug behavior and ensures the best possible fit with a generalized model.}

    \label{fig:workflow}
\end{figure}

A comprehensive illustration, presented in Figure \ref{fig:workflow}, elucidates the architectural underpinnings of our innovative approach to generalizing compartmental modeling. The following is a description of the organization of this paper: Section 2 is devoted to the presentation of the methodology employed in this study. Here, we delineate our proposed generalized modeling approach and illustrate the use of physics-informed neural networks for inferring model parameters from real data. Section 3 presents the four original PK and PK-PD compartmental models selected for examination. It also describes the specific modifications each model requires for simplification using our novel modeling approach. Section 4 presents the findings obtained through the application of these two neural network-based solvers and the two novel modeling approaches. Finally, section 5 provides a summary and a discussion of our findings and possible limitations.

\section{Materials and Methods}\label{sec:models}
\subsection{Proposed Modeling Concept}\label{sec:theory}
In Pharmacokinetics modeling, "compartments" are abstract mathematical constructs designed to simplify the representation of drug distribution within the human body. These compartments do not correspond to specific anatomical or physiological spaces; rather, they serve as valuable tools for modeling the behavior of substances introduced into the body \cite{talevi2024compartmental}. Multi-compartment models enable the description of drug plasma concentration as decaying in multiple exponential phases, thereby capturing the complexity of distribution and elimination processes. The selection between compartmental and non-compartmental models is typically guided by the objectives of the analysis. Compartmental models are frequently employed to explore Pharmacokinetics variability across different populations or conditions, and to inform dosage adjustments accordingly. 

Compartmental models, being conceptual rather than physical constructs, require careful selection to accurately represent the underlying data. The selection process is often guided by the number of exponential terms identified in the logarithmic plots of concentration-time or effect-time profiles. However, determining the correct number of compartments can be particularly challenging, especially in cases of sparse data. Sparse data, often resulting from infrequent sampling, can obscure the detection of distinct phases of drug distribution and elimination. This may lead to an oversimplified interpretation, such as incorrectly classifying a multi-compartment system as a single-compartment model. Additionally, non-linear kinetics, where drug concentration over time deviates from a simple exponential decay, further complicates the analysis of logarithmic plots and the accurate determination of the number of compartments \cite{Mukherjee2022}. Moreover, most Pharmacokinetics data are derived from the central compartment (e.g., plasma), which can lead to parameter identifiability issues as the number of compartments increases. Adding more compartments raises the risk of different parameter sets producing similar fits to the data, complicating the identification of unique parameter values. High variability or noise in the data can also obscure the distinct phases, contributing to potential misidentification of compartments. Other complicating factors include the complexity of drug behavior, such as the overlapping distribution and elimination phases. In such cases, traditional models may prove inadequate, necessitating the adoption of more sophisticated approaches or the inclusion of additional compartments. The assumptions that underpin compartmental models, such as the homogeneity of compartments, may not be universally applicable, thereby further complicating the process of identifying the relevant compartments.

In light of these challenges, we investigate how distinct drugs can manifest disparate Pharmacokinetics profiles while being modeled using a unified two-compartment framework, irrespective of whether conventional modeling methodologies categorize them as one-compartment, two-compartment, or multi-compartment systems. In particular, our objective is to ascertain how discrepancies in drug absorption rates can be identified when they are described by the same system of two ODEs. The objective of this study is to develop a generalized model and apply it to various Pharmacokinetics phenomena with the aim of challenging and revising the traditional modeling paradigm. We propose that all multi-compartment Pharmacokinetics models can be simplified into two compartments through one of two modifications: either by incorporating drug trapping phenomena using fractional derivative models or by considering the absorption rate as time-varying. These approaches have the potential to generalize modeling methodologies and provide novel insights into variations in absorption rates or memory effects across different drugs' concentration-time profiles. The key parameters driving changes in the slope of the curve are the absorption rate and the fractional derivative order, allowing for different exponential decay behaviors to reflect either time-varying parameters or drug trapping phenomena across various models and drugs.

In a PK-PD model, incorporating time-varying parameters provides valuable insights into phenomena such as drug tolerance, chemotherapy resistance, and persistence in the use of anticancer agents. The present study explores these concepts by analyzing the dynamics of these parameters over time, offering a deeper understanding of their impact on drug efficacy. These techniques enable the creation of simpler models while preserving the flexibility needed to accurately describe the various phases of drug concentration decay and efficacy over time.
\subsection{Methodology}\label{sec:methodology}

Improving the accuracy of parameter estimation in PK/PK-PD modeling and uncovering new insights in systems described by ordinary differential equations calls for a method that can (i) account for time-varying parameters to more effectively model drug absorption dynamics in PK frameworks and drug effects in PK-PD models, thereby facilitating the simulation of drug tolerance and cancer drug resistance, and (ii) utilize fractional calculus to represent distributed and delayed drug responses in PK models or cancer cell death dynamics in integrated PK-PD models. To achieve these goals, we implement Physics-Informed Neural Networks (PINNs), a widely acknowledged approach in scientific machine learning, to address both forward and inverse problems linked to differential equations. Through the use of a deep fully connected neural network and physics-informed loss functions, we can tackle the inverse problem in our novel modeling approach, which integrates time-varying, constant, piecewise constant, and fractional components. For handling fractional-order differential equations, we employ fPINNs, a variation of PINNs specifically designed for such equations. The specifics of these methods are detailed in the following methodology subsections.

\subsection{PINNs with Time-varying parameters}

Building upon the concept of Physics-Informed Neural Networks (PINNs), we propose a deep learning framework that incorporates the differential equations governing compartmental Pharmacokinetics and Pharmacodynamics models. In this framework, a neural network parameterized by \(\theta\) takes time \(t\) as input and produces an output vector representing the state variables \(\hat{u}(t; \theta) = (\hat{u}_1(t; \theta), \hat{u}_2(t; \theta))\), which approximates the solution to the ordinary differential equations \(u(t)\). To solve the inverse problem, where both time-varying and potentially constant or piecewise constant parameters are unknown, we extend the framework by introducing an additional output that approximates the time-dependent unknown parameter.

\begin{figure}[!ht]
    \centering
    \includegraphics[width=\linewidth]{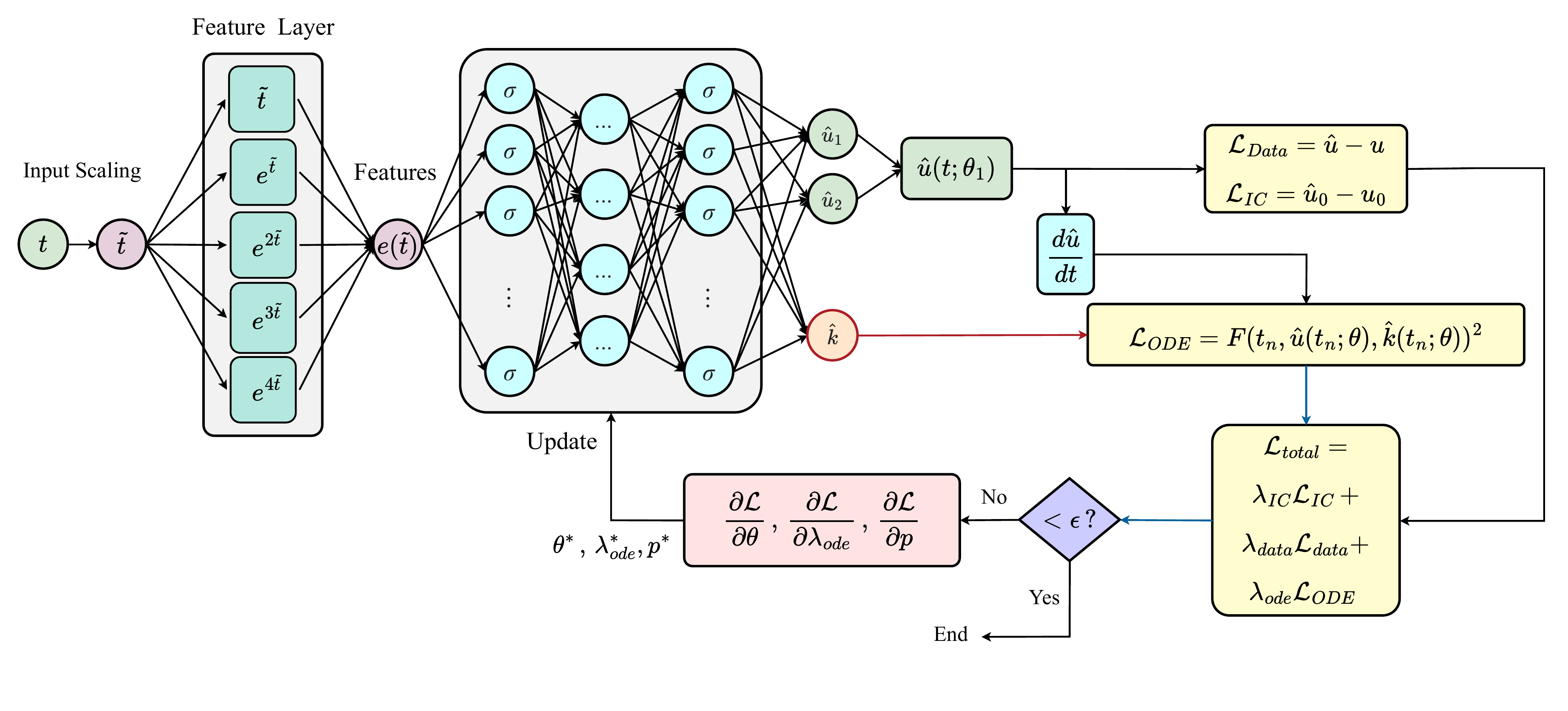}
    \caption{\textbf{PINNs block in CMINNs method.} The physics-informed neural network model starts from an input scaling layer that normalizes all values to the order of 1. If the model is complicated we use the feature layer to add exponential terms within the neural networks. Then, as the output of neural network model we obtain the solution of the system of ODEs along with the time-varying parameter(s). The output goes to an optimization block where it minimizes the physical loss as well as the data loss and updates the neural network parameters. We employ adaptive weights for the physics loss, where the coefficient $\lambda_{ode}$ is updated simultaneously with the neural network parameters $\theta$ and the compartment model's constant parameters $p$.}
    \label{fig:pinns}
\end{figure}

The architecture of the PINNs is depicted in Figure \ref{fig:pinns}. A key step in this process is constraining the neural network to satisfy both the observed data points of \(u(t)\) and the system of ODEs. This is accomplished by constructing a loss function that incorporates terms corresponding to both the observational data and the compartmental model. Specifically, let us assume we have measurements \(u_{\text{data}} = \{u_1, u_2, \ldots, u_M\}\) at various time points \(t_1, t_2, \ldots, t_{M_{\text{}}}\). The neural network must satisfy the ODE system at specific time instances \(t_1, t_2, \ldots, t_{N_{\text{}}}\), where these time points may be non-uniform and can be selected arbitrarily. Here, \(N\) represents the number of collocation points, and \(M\) represents the number of data points.

For computing the total loss, we utilize the Self-Adaptive Loss Balanced method \cite{xzp2022,mcclenny2023selfadaptive}. The total loss function is defined as a function of \(\theta, p, \lambda_{\text{ode}}\), where \(p\) denotes the unknown parameters of the ODEs, and \(\lambda_{\text{ode}}\) is a vector representing the individual loss weights for the state variables, i.e., \(\lambda_{\text{ode}} = (\lambda_{1}, \lambda_{2})\). We note that \(\lambda_{\text{data}}\) and \(\lambda_{\text{IC}}\) are constant values, set to 0 during the first stage of training and 1 in the second stage of training, and are not trainable variables in our neural network. The total loss is computed as follows:

\begin{equation}
    \mathcal{L}(\theta, p, \lambda_\text{ode}) = \lambda_\text{IC}\mathcal{L_{\text{IC}}}(\theta) + \lambda_\text{data}\mathcal{L_{\text{data}}}(\theta) + \lambda_\text{ode}\mathcal{L_{\text{ode}}}(\theta, p), 
\end{equation}

where
\begin{equation}
    \mathcal{L_{\text{IC}}}(\theta) = \left(u(t_0) - \hat{u}(t_0; \theta)\right)^2
\end{equation}

\begin{equation}
    \mathcal{L_{\text{}}}(\theta) = \frac{1}{M_{\text{}}} \sum_{m=1}^{M_{\text{}}} (u(t_m) - \hat{u}(t_m; \theta))^2
\end{equation}

\begin{equation}
    \mathcal{L_{\text{ode}}}(\theta, p) = \frac{1}{N_{\text{}}}\sum_{n=1}^{N_{\text{}}} \left(\frac{d\hat{u}}{dt}\bigg|_{t_n} - F(t_n,\hat{u}(t_n; \theta), k(t_n; \theta) ; p)\right)^2.
\end{equation}\\

We note that both the loss function derived from the data, denoted by $L_{data}$, and the loss function derived from the initial condition, denoted by $L_{IC}$, quantify the difference between the neural network outputs and the observed data. As a result, they can be categorised as supervised loss functions. On the other hand, $L_{\text{ode}}$, which is derived from the underlying ODE system, functions as an unsupervised loss. In the final optimization stage, we jointly estimate the neural network parameters, $\theta^*$, along with the unknown ODE parameters, $p^*$, where $p^*$ represents constant or piecewise constant parameters, and $k(t_n; \theta)$ represents time-varying parameters, by minimizing the overall loss function. This is achieved through gradient-based optimization techniques, such as the Adam optimizer \cite{kingma2014adam}. Moreover, the vector $\lambda^*_{\text{ode}}$ is obtained by adjusting the adaptive weights during each epoch, according to:
\begin{equation}
    \theta^*, p^*, \lambda^*_\text{ode} = \arg \max_{\lambda_\text{ode}} \min_{\theta, p} \mathcal{L}(\theta, p, \lambda_\text{ode}). 
\end{equation}

This represents a min-max optimization problem where we aim to minimize the loss by updating the compartment model and neural network parameters, while simultaneously maximizing the weight associated with the physics-based loss. For the training process, where our goal is to simultaneously predict the unknown function $k(t;\theta)$ and estimate the constant parameter values, we use the Adam optimizer with default hyperparameters and a different learning rate. Training is conducted either on the entire time domain for PK models or on sequential intervals for PD model. Since our total loss comprises two supervised losses and one unsupervised loss, we adopt a two-stage training strategy for PK compartment models as follows:

\begin{enumerate}
  \item  We have observed that supervised training typically yields a faster convergence than unsupervised training. Consequently, the network is initially trained using the two supervised losses, namely, the data loss function, denoted as $\mathcal{L}_{data}$, and the initial condition loss function, denoted as $\mathcal{L}_{IC}$. This is done for a specified number of iterations. This preliminary training phase allows the network to rapidly align itself with the observed data points.
  \item Subsequently, the training process is continued, with all three losses being incorporated.
\end{enumerate}
Empirical observations demonstrate that this two-stage training approach expedites network convergence \cite{Ahmadi2024}. By initially warming up the neural network with parameters that produce an output fitted to the data, we start the full training with a significantly smaller loss, which accelerates convergence.
We did not use this method for the PD model because it is trained in sequential intervals. In each interval, we may not have any data, so the first stage is unnecessary for this type of training. Another significant point is that in the time-domain decomposition method, we observe the data over the entire time range, and only the training on the physics law is done in sequential intervals. This allows us to add an additional equation to further constrain the physics component, which we will explain in the next section. The specific PINNs parameter setup for each model is detailed in Table \ref{table:pinns_param}.

When applying PINNs to compartmental models, several challenges arise. One significant issue is the minimal changes in drug concentration for certain drugs, such as amiodarone, which cause the neural network's output to rapidly approach zero. To address this, we propose adding a feature layer with exponential terms. Another challenge is the typically large time range, as we often examine the effect or concentration over extended periods. This necessitates the inclusion of auxiliary layers in the PINNs to normalize or scale the inputs, bringing all values to the same order of magnitude, as demonstrated in Figure \ref{fig:pinns}. Here, we added an input scaling layer and normalized all equations to the range [0,1] to ensure the outputs are within the same range as the scaled inputs.

For PK-PD models, we encounter additional challenges due to the presence of short-term spikes in the PD models, resulting from repeated drug administration. These spikes can cause the PD model solutions to exhibit multiple peaks. A primary challenge is the issue of catastrophic forgetting in the neural network when trained over large time ranges, leading to a failure to capture one or more peaks and sudden spikes in a multi-dose administration schedule. To mitigate this, we propose a sequential learning approach for PINNs, where the model is trained over smaller time intervals, with a focus on expected spike times corresponding to the drug administration schedule. This method offers two key benefits: first, it helps overcome the problem of catastrophic forgetting; second, it allows us to infer the value of constant parameters, such as the index of drug efficacy, within each interval. By treating these parameters as piecewise constants, we can observe how they change following multi-dose drug administration, a topic we will explore further in the discussion of the fourth model.

\subsection{fPINNs}
Fractional Physics-Informed Neural Networks (fPINNs) are an extension of PINNs, specifically designed to address both fractional partial and ordinary differential equations. Herein, we construct fractional PK and PK-PD models of the same type with different manifestations by adopting various modeling approaches. Subsequently, we obtain the parameters of the fractional models by solving the inverse problem using the fPINNs algorithm across the different modeling approaches and compare the results with those obtained using other proposed methods. This holistic approach enables us to to select the most suitable fractional model, which is representative of the anomalous diffusion and absorption laws of the drug concentration.

One of the key features of fractional models is their non-local or memory property, which distinguishes them from integer-order differential operators. This property implies that the future state of a model depends not only on its current state but also on its entire history. For an in-depth exploration of the definitions and theoretical foundations of fractional calculus, we refer the reader to  \cite{podlubny1999fractional,kilbas2006theory,khalil2014new}.
 In this work, we employ the Caputo fractional derivative, which requires an initial condition involving the state variable.

\textbf{Definition:} The Caputo fractional derivative operator $ \prescript{C}{0}{D_x^\alpha} $ of order $\alpha $ is defined as:

\begin{equation}
\prescript{C}{0}{D_x^\alpha} f(x) = \frac{1}{\Gamma(m - \alpha)} \int_{0}^{x} \frac{f^{(m)}(t)}{(x - t)^{\alpha - m + 1}} \, dt, \quad \alpha > 0,
\end{equation}

where  $m - 1 < \alpha \leq m $, $ m \in \mathbb{N} $, and $ x > 0 $.

The Caputo fractional derivative operator is a linear operation, analogous to the integer-order derivative:

\begin{equation}
\prescript{C}{0}{D_x^\alpha} \left(\lambda f(x) + \mu g(x)\right) = \lambda \prescript{C}{0}{D_x^\alpha} f(x) + \mu \prescript{C}{0}{D_x^\alpha} g(x),
\end{equation}

where $ \lambda $ and $ \mu $ are constant values. We note that the Caputo differential operator reduces to the classical derivative when $ \alpha $ is an integer. The initial conditions for fractional differential equations using the Caputo derivative are analogous to those for integer-order differential equations, making this definition particularly suitable for modeling a wide range of physical processes.

It is well known that in machine learning, the design of the loss function and the order of computation are crucial for optimizing neural networks. In the context of parameter estimation in fPINNs, fitting the actual data should take precedence over fitting the mathematical model constructed under ideal conditions. Therefore, we restructured the iterative fitting process of the fPINN method into a two-stage training process. Initially, the loss function is configured to include only the label penalty term for the real data, i.e., the supervised losses. Once the model accurately fits the data, the equation penalty term (unsupervised loss) is introduced into the loss function with a smaller penalty weight($1e^{-4}$). The non-local nature of the fractional derivative operators in Equation (6) introduces significant challenges in solving it. The $L_{IC}$ and $L_{data}$ will be computed in the same way as in the PINNs method, while we have a different formula for $L_{ode}$, where we use the Caputo derivative definition instead of the integer-order derivative of the equation, as follows:
\begin{equation}
    \mathcal{L_{\text{ode}}}(\theta, p) = \frac{1}{N^{\text{ode}}}\sum_{n=1}^{N^{\text{ode}}} \left(\prescript{C}{0}{D_t^\alpha}\hat{u}\bigg|_{t_n} - F(t_n,\hat{u}(t_n; \theta), k(t_n; \theta) ; p)\right)^2.
\end{equation}\\
Typically, approximating fractional derivative operators involves a discrete convolution form (see, e.g., \cite{26,27,28,29,30}) for $0 \leq j \leq n$, where $0 < n \leq K$. The fractional derivative can be expressed as:
\begin{equation}
\prescript{C}{0}{D_t^\alpha} u(t_n) = \sum_{j=0}^{n} \omega^{(\alpha)}_{n-j} (u^j - u^0)
\end{equation}
where $\tau$ is the time step size, $K$ is defined as $K = \frac{T}{\tau}$, and $\omega^{(\alpha)}_{n-j}$ are the convolution quadrature weights. Implementing this convolution directly requires $\mathcal{O}(K)$ active memory and $\mathcal{O}(K^2)$ arithmetic operations, making the computation quite expensive. 
To reduce the memory requirements and computational costs associated with the discrete convolution used for approximating fractional derivatives, several methods have been proposed \cite{31,32,33,34,35,36}. In general, fractional differential operators are often discretized using numerical methods such as the finite difference method (FDM), finite element method (FEM), or spectral method (SM). In this paper, we employ the second-order fractional backward differentiation formula (FBDF) algorithm \cite{fbdfl} within the FDM framework. In the second-order FBDF Extended scalar auxiliary variable(SAV) Method Preserving Energy Dissipation, the time interval $[0, T]$ can be divided into $K$ equal subintervals with a time-step size $\tau = \frac{T}{K}$. Let $t_n = n\tau$ and $u^n = u(x, t_n)$ for $0 \leq n \leq K$. We use the following notations:
\begin{equation}
D_\tau^1 u^{n+1} = \frac{3u^{n+1} - 4u^n + u^{n-1}}{2\tau}, \quad u^{n+1} = 2u^n - u^{n-1}.
\end{equation}
Using the FBDF method \cite{fbdfl}, the Caputo derivative can be discretized as
\begin{equation}
\left[ \prescript{C}{0}{D_t^\alpha} u \right]_{t = t_n} = D_\tau^\alpha u^n + R^n
,  \quad R^n = \mathcal{O}(\tau^2)
\end{equation}
where
\begin{equation}
D_\tau^\alpha u^n = \frac{1}{\tau^\alpha} \sum_{j=0}^{n} \omega^{(\alpha)}_{n-j} (u^j - u^0).
\end{equation}
Here, $\{\omega^{(\alpha)}_n\}$ are the coefficients of the Taylor expansions of the generating function:
\begin{equation}
\omega^{(\alpha)}(z) = \left(\frac{3}{2} - 2z + \frac{1}{2}z^2\right)^\alpha = \sum_{n=0}^{\infty} \omega^{(\alpha)}_n z^n.
\end{equation}


\begin{figure}[!ht]
    \centering
    \includegraphics[width=\linewidth]{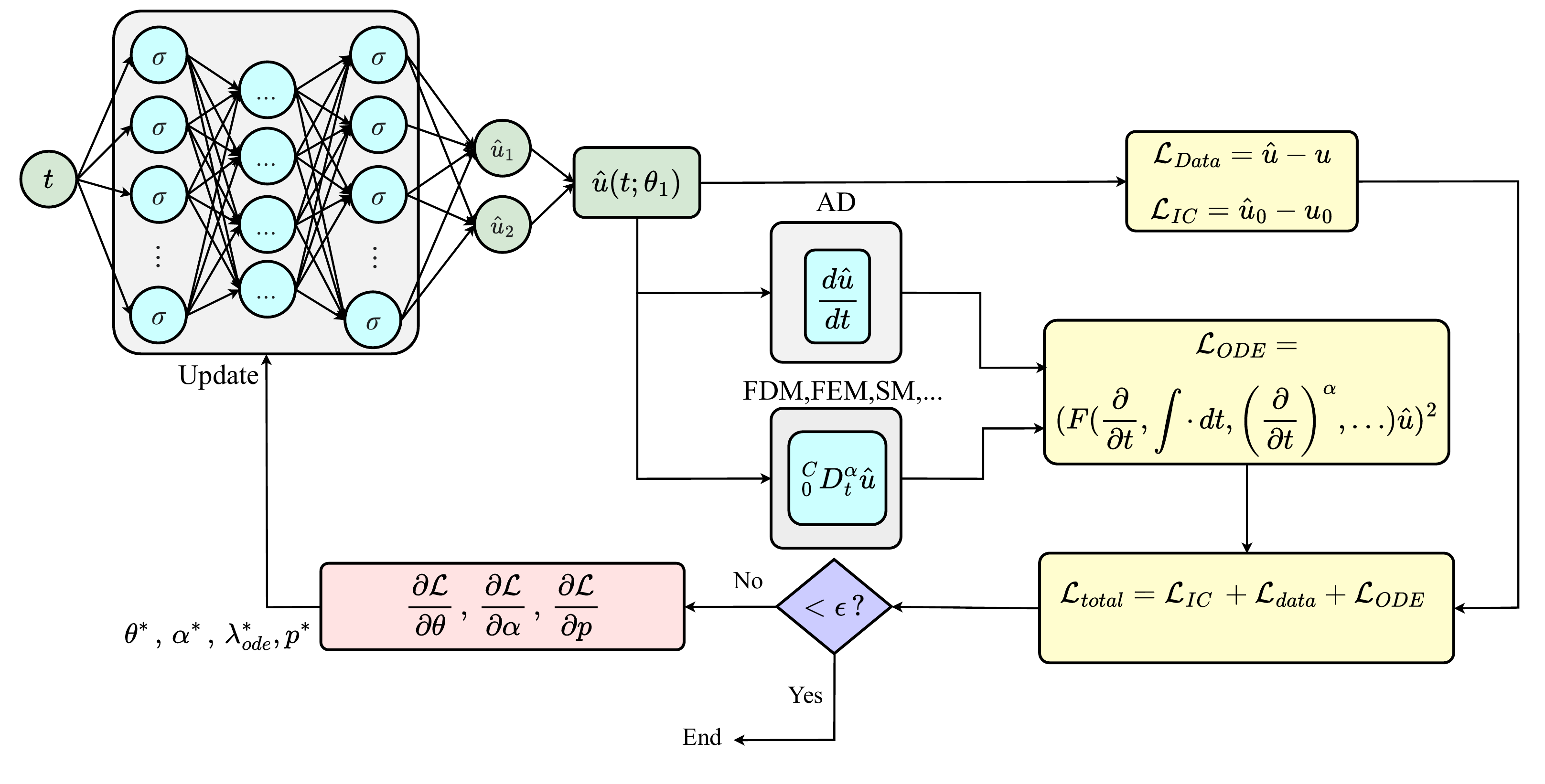}
    \caption{\textbf{fPINNs Block in CMINNs method.} The abbreviation “AD” refer to automatic differentiation. In this work, we use the Finite Difference Method (FDM) and FBDF for discretization of fractional order derivative to compute numerical solutions. In this model, the neural network output is passed to the AD block for integer-order derivative computation and to the FDM block for fractional-order differentiation. The computed derivatives are then incorporated into the equation to calculate the physics loss and simultaneously update the fractional compartment model's constant parameters $p$ and $\alpha$, along with the neural network parameters $\theta$.}
    \label{fig:fpinns}
\end{figure}
%
\section{Mathematical Models}

In this section, we present four mathematical models, three of which are focused on pharmacokinetics for three distinct drugs. These models detail the dynamics of drug concentration in the blood. The fourth model combines pharmacokinetics and pharmacodynamics, with an emphasis on the pharmacodynamic aspects. A pharmacokinetic model typically quantifies the changes in drug concentration in the bloodstream over time, whereas a 
pharmacodynamics model focuses on the relationship between drug concentration at the site of action and the resultant biological effect. These models are meticulously crafted to capture the dynamic interactions within specific biological processes, such as drug absorption. They provide a physics-based understanding of the behavior and characteristics of the systems under examination. This approach enables a more detailed simulation of the drug's behavior within the body, thereby facilitating the development of personalized treatment strategies based on the dynamic responses observed.
\subsection{Model 1: Two-Compartment Pharmacokinetic Model for Gentamicin}
Managing chronic wounds is a significant area of focus in medical and biomaterials research, especially given the rising concerns related to multi-drug resistant microbes and antibiotic-resistant bacteria. To tackle these issues, a poly(vinyl alcohol)/gentamicin (PVA/Gent) hydrogel has been developed for wound treatment. This hydrogel is designed to deliver gentamicin locally over an extended period, eliminating the need for systemic drug administration.

A two-compartment pharmacokinetic model is utilized to represent the release of gentamicin from the PVA/Gent hydrogel, which is specifically aimed at wound dressings for treating severe chronic wounds. This model omits an elimination rate because the drug is not present in the bloodstream where it would be cleared. The model is based on the data described in \cite{miskovic2023two}, which outlines the preparation and analysis of the hydrogel system. The PVA/Gent hydrogel was created by physically cross-linking a poly(vinyl alcohol) solution using a freeze-thaw method, followed by soaking in a gentamicin solution at 37°C for 48 hours. The amount of gentamicin released was measured using high-performance liquid chromatography combined with mass spectrometry. This methodology addresses the challenge of controlling gentamicin release from the hydrogel, which is crucial for effective chronic wound care and healing.

Considering the classical two-compartment pharmacokinetic model, the equations governing the system are as follows:

\begin{equation}
\begin{aligned}
    \frac{dA_1}{dt} &= \frac{-k}{V_1} A_1(t) + \frac{k}{V_2} A_2(t) + f_1(t),\\
    \frac{dA_2}{dt} &= \frac{k}{V_1} A_1(t) - \frac{k}{V_2} A_2(t), 
    \label{eq:gen12}
\end{aligned}
\end{equation}
where $A_i$ $(i = 1, 2)$ represents the mass of gentamicin in compartments $1$ and $2$, respectively, and $V_i$ $(i = 1, 2)$ denotes the volume of these compartments. Based on previous findings, we set $V_1 = 254.5 \, \text{mm}^3$ and $V_2 = 1000 \, \text{mm}^3$. Additionally, $f_1$ represents the supply of drug to compartment $1$. The dimensions of $k$ is [mm$^3$/day].

We adapt this model according to the CMINNs modeling approach as follows:

\begin{enumerate}
    \item When utilizing PINNs, the constant value $k$ must be modified to be a time-dependent function. Consequently, the equations are rewritten as:
    \begin{equation}
    \begin{aligned}
    \label{eq:pinn_general}
        \frac{dA_1(t)}{dt} &= -k(t) \left(\frac{A_1(t)}{V_1} - \frac{A_2(t)}{V_2}\right) + f_1(t), \\
        \frac{dA_2(t)}{dt} &= k(t) \left(\frac{A_1(t)}{V_1} - \frac{A_2(t)}{V_2}\right).
    \end{aligned}
    \end{equation}

    \item In the context of fPINNs, it is necessary to introduce a fractional derivative into the model. Thus, the system of ODEs defined in \eqref{eq:gen12} is transformed into the following commensurate fractional PK model of equal-order in the Caputo sense:
    \begin{equation}
    \begin{aligned}
    \label{eq:fgeneral}
        \,^{C}_{0}D^{\alpha}_{t} A_1 &= -k\left(\frac{A_1(t)}{V_1} - \frac{A_2(t)}{V_2}\right) + f_1(t),\\
        \,^{C}_{0}D^{\alpha}_{t} A_2 &= k\left(\frac{A_1(t)}{V_1} - \frac{A_2(t)}{V_2}\right),  \quad 0 < \alpha < 1, 
    \end{aligned}
    \end{equation}
    
\end{enumerate}
where \( \,^{C}_{0}D^{\alpha}_{t} \) denotes the Caputo fractional derivative of order \( \alpha \).
To satisfy mass conservation, the derivative orders for both compartments must be equal. If they differ, the system becomes non-commensurate, leading to significant issues. Non-commensurate systems exhibit properties that violate mass balance, particularly concerning the consistency of rate units. For example, if the mass flux leaving one compartment is defined by a non-integer order rate, it would enter another compartment with a different non-integer order rate, thereby violating mass balance \cite{Dokoumetzidis2010}.


\subsection{Model 2: Two-Compartment Pharmacokinetic Model for Amiodarone}\label{sec:magin}

The pharmacokinetics of a single dose of amiodarone, an antiarrhythmic drug known for its non-exponential pharmacokinetics and significant clinical implications due to drug accumulation during long-term administration, have been analyzed using a general two-compartment model. The model was applied to a dataset from intravenous administration studies on volunteer subjects given 400 mg doses. This dataset was initially presented in \cite{holt1983amiodarone}. 

The governing equations of this model are described by the following system of linear ordinary differential equations:
\begin{equation}
\begin{aligned}
    \frac{dA_1(t)}{dt} &= -k_{12} A_1(t) + k_{21} A_2(t) - k_{10} A_1(t) + I_1(t), \\
    \frac{dA_2(t)}{dt} &= k_{12} A_1(t) - k_{21} A_2(t).
\end{aligned}
\end{equation}

Here, $A_1(t)$ and $A_2(t)$ represent the mass or molar amounts of Amiodarone in the respective compartments. The parameters  $k_{ij}$ govern the mass transfer between the compartments and the elimination from each, where the notation $k_{12}$ denotes the rate of transfer from compartment 1 to compartment 2, and $k_{10}$ denotes elimination from compartment 1, and so forth. The units for all $k_{ij}$ rate constants are $(1/\text{time})$. The input rates $I_1(t)$ representing the drug dose and administration schedule and it may be zero, constant, or time-dependent.

Amiodarone displays unusual pharmacokinetic characteristics that conventional models fail to adequately capture. The long-term pharmacokinetics of this highly cationic  amphiphilic drug presents challenges that traditional pharmacokinetic concepts cannot address. To evaluate the effectiveness of our proposed CMINNs model, we apply it to a scenario known for its anomalous pharmacokinetic behaviors. The model is reformulated as follows:

\begin{enumerate}
    \item Utilizing PINNs, we modify the absorption rate constant \(k_{12}\) to a time-dependent function \(k_{12}(t)\), which allows us to capture the natural fluctuations in the absorption rate, while keeping \(k_{21}\) and \(k_{10}\) as constant values:

    \begin{equation}
    \begin{aligned}
    \label{eq:pinns_magin}
        \frac{dA_1(t)}{dt} &= -k_{12}(t) A_1(t) + k_{21} A_2(t) - k_{10} A_1(t), \\
        \frac{dA_2(t)}{dt} &= k_{12}(t) A_1(t) - k_{21} A_2(t). 
    \end{aligned}
    \end{equation}

    \item Based on the fractional model proposed in \cite{Dokoumetzidis2010}, which outlines a generalized approach for the fractionalization of compartmental models by mixing different fractional orders, we consider a two-compartment fractional pharmacokinetic model. In this model, Compartment 1 (central) represents general circulation and well-perfused tissues, while Compartment 2 (peripheral) corresponds to deeper tissues. The system considers three transfer processes (fluxes): elimination from the central compartment and a mass flux from the central to the peripheral compartment, both of which are assumed to follow classic first-order kinetics. However, the flux from the peripheral to the central compartment is modeled with slower fractional kinetics to account for tissue trapping.

    This system is mathematically formulated as follows:

    \begin{equation}
    \begin{aligned}
    \label{eq:fmagin}
        \frac{dA_1(t)}{dt} &= -(k_{12}(t) + k_{10}) A_1(t) + k_{21} \,^{C}_{0}D^{1-\alpha}_{t} A_2(t), \\
        \frac{dA_2(t)}{dt} &= k_{12}(t) A_1(t) - k_{21} \,^{C}_{0}D^{1-\alpha}_{t} A_2(t),  \quad 0 < \alpha < 1.
    \end{aligned}
    \end{equation}

\end{enumerate}


\subsection{Model 3: Three-Compartment Pharmacokinetics Model for Talaporfin Sodium }

Talaporfin sodium (2.5 mg/kg) was intravenously administered to an adult male dog. The concentration of talaporfin sodium in plasma was quantified using a spectrophotometer. Additionally, the fluorescence of talaporfin sodium in skin and myocardium was measured in vivo with a specially developed system. The skin fluorescence, excited by 409 ± 16 nm light to match the talaporfin sodium Soret band, and the myocardial fluorescence, excited by 663 ± 2 nm light to match the talaporfin sodium Q band, were both quantified using a spectrometer. We use the data and the model described in \cite{bb0} in our study. This model estimates the interstitial concentration of talaporfin sodium based on changes in plasma concentration and myocardial fluorescence, enhancing our understanding of its efficacy interstitial photodynamic therapy (PDT). It also integrates differential rate equations across the plasma, interstitial space, and cellular compartments, incorporating specific compartment volumes, concentrations, and rate constants.

The dynamics of talaporfin sodium across these compartments is governed by the following differential rate equations:
\begin{equation}
\begin{aligned}
    V_1 \frac{dC_1}{dt} &= -(k_{10} + k_{12}) V_1 C_1 + k_{21} V_2 C_2, \\
    V_2 \frac{dC_2}{dt} &= k_{12} V_1 C_1 - (k_{21} + k_{23}) V_2 C_2 + k_{32} V_3 C_3, \\
    V_3 \frac{dC_3}{dt} &= k_{23} V_2 C_2 - k_{32} V_3 C_3,
\end{aligned}
\end{equation}
where $V_i$ and $C_i$ represent the volume and concentration in the $i^{th}$ compartment (i = 1, 2, 3), respectively. The rate parameters  $k_{10}, k_{12}, k_{21}, k_{23}, k_{32}$ regulate the transfer of mass between compartments and the elimination from each. The volumes $[v_1, v_2, v_3]$ are set to $[394, 251, 970]$ mL, as determined by histological assessments reported in \cite{TalaporfinInterstitial}.

According to the CMINNs workflow illustrated in Figure \ref{fig:workflow}, we need to reduce the model to two compartments. However, since this model includes data for all compartments—an uncommon feature in compartmental models—it is challenging to achieve a good fit with traditional methods. Therefore, our first step is to rewrite the model using a time-varying absorption rate and fractional derivatives to enhance its flexibility in capturing changes in the concentration-time profiles across all compartments. Thus, in \ref{sec:3comp}, we reformulate the three-compartment model, and in \ref{sec:3to2}, we proceed to the compartment reduction block of CMINNs to reformulate the PK model with two compartments.

\subsubsection{Reformulation of the Three-Compartment Model with Time-Varying Absorption Rates and Fractional Derivatives}\label{sec:3comp}

To address the dynamic nature of drug absorption rates, we assume that $k_{12}$, the primary rate constant, is a function of time. Additionally, by incorporating fractional calculus, we aim to better capture the delay in drug response. We will compare the results of two different methods for fractionalizing the model: applying fractional derivatives either to the right-hand side or the left-hand side of the equations, as introduced in Equations \eqref{eq:fgeneral} and \eqref{eq:fmagin}. Using PINNs with a time-varying absorption rate, in the three-compartment model, we only need to replace $k_{12}$ with $k_{12}(t)$, which is one of the outputs of our PINNs model.
To change the model to include fractional derivatives, there are two approaches explained in detail in reference  \cite{Borkor2023}. The paper presents the following two fractional models formulated from a classical pharmacokinetics compartmental system: commensurable and implicit non-commensurable models.
\begin{itemize}
    \item[a)] The commensurate fractional three-compartment pharmacokinetic model is described by the following set of fractional differential equations:
    \begin{equation}
    \begin{aligned}
    \label{eq:3comp_fpinn_a}
    V_1 \,^{C}_{0}D^{\alpha}_{t} C_1 &= -(k_{10} + k_{12})V_1 C_1 + k_{21} V_2 C_2, \\
    V_2 \,^{C}_{0}D^{\alpha}_{t} C_2 &= k_{12} V_1 C_1 - (k_{21} + k_{23}) V_2 C_2 + k_{32} V_3 C_3, \\
    V_3 \,^{C}_{0}D^{\alpha}_{t} C_3 &= k_{23} V_2 C_2 - k_{32} V_3 C_3.
    \end{aligned}
    \end{equation}

    \item[b)] The non-commensurate fractional three-compartment PK model introduces a different fractional order in one of the compartments. The governing  equations are written as follows:
    \begin{equation}
    \begin{aligned}
    \label{eq:3comp_fpinn_b}
        V_1 \frac{dC_1}{dt} &= -(k_{10} + k_{12})V_1 C_1 + k_{21}V_2 C_2, \\
        V_2 \frac{dC_2}{dt} &= k_{12}V_1 C_1 - (k_{21} + k_{23})V_2 C_2 + k_{32}V_3 \,^{C}_{0}D^{1-\alpha}_{t} C_3, \\
        V_3 \frac{dC_3}{dt} &= k_{23}V_2 C_2 - k_{32}V_3 \,^{C}_{0}D^{1-\alpha}_{t} C_3.
    \end{aligned}
    \end{equation}
\end{itemize}

We need to infer the fractional order $\alpha$ along with the constant values of the unknown parameters $k_{10}$, $k_{12}$, $k_{21}$, $k_{23}$, and $k_{32}$.
\subsubsection{Reduction of the Three-Compartment Model to a Two-Compartment Model Using CMINNs}\label{sec:3to2}

In this section, we reformulate the Three-Compartment Model to a Two-Compartment Model with a time-varying absorption rate and fractional order derivative as follows:

\begin{enumerate}
    \item Generalized model with time-dependent parameter:
    \begin{equation}
    \begin{aligned}
        V_1 \frac{dC_1}{dt} &= -(k_{10} + k_{12}(t))V_1 C_1 + k_{21} V_2 C_2, \\
        V_2 \frac{dC_2}{dt} &= k_{12}(t) V_1 C_1 - k_{21}V_2 C_2.
    \end{aligned}
    \label{eq:generalized_model}
    \end{equation}
    
    \item Fractional generalized model:
    \begin{equation}
    \begin{aligned}
        V_1 \frac{dC_1}{dt} &= -(k_{10} + k_{12})V_1 C_1 + k_{21}V_2 \,^{C}_{0}D^{1-\alpha}_{t}C_2, \\
        V_2 \frac{dC_2}{dt} &= k_{12}V_1 C_1 - k_{21}V_2 \,^{C}_{0}D^{1-\alpha}_{t}C_2.
    \end{aligned}
    \label{eq:fractional_model}
    \end{equation}
\end{enumerate}


\subsection{ Model 4: Pharmacokinetic-Pharmacodynamic Model for Tumor Growth and Treatment in Animal}\label{sec:pkpd}

A pharmacokinetic-pharmacodynamic (PK-PD) model was developed to simplify the complex behavior of tumor growth, making it more suitable for use in preclinical oncology drug development \cite{simeoni2004predictive}. This model, described through a system of ODEs, connects the dosage of a drug to tumor growth patterns observed in animal experiments. It separates the tumor growth dynamics into two categories: untreated and treated groups. Tumors in the untreated group initially follow an exponential growth phase, which transitions into a linear phase. In contrast, in the treated group, the tumor growth rate is smaller in relation to both the drug concentration and the number of actively proliferating tumor cells.
The differential equation governing the unperturbed tumor growth $w(t)$  is given by:
\begin{align}
\label{eq:tumor}
\frac{dw(t)}{dt} = \frac{\lambda_0 \cdot w(t)}{[1 + \left(\frac{\lambda_0}{\lambda_1} \cdot w(t)\right)^{\Psi}]^{1/\Psi}} ,  
w(0) = w_0.
\end{align}

When $\Psi$ becomes large enough, Equation \eqref{eq:tumor} accurately models the transition in tumor growth dynamics. In particular, for tumor mass $w(t)$ below a critical threshold $w_{\text{th}}$, the term $\frac{\lambda_0}{\lambda_1} \cdot w(t)$ in the denominator is small compared to 1, resulting in a growth rate primarily dominated by $\lambda_0 \cdot w(t)$, representing exponential growth. On the other hand, when $w(t)$ surpasses $w_{\text{th}}$, the 1 in the denominator becomes negligible, shifting the growth rate to a linear behavior driven by $\lambda_1$. In practice, setting $\Psi$ to 20 or higher ensures a distinct transition from exponential (first-order) to linear (zero-order) growth, effectively replicating the sharp change seen in the original switching systems-based model.

This model introduces an innovative transit compartment system to simulate the delayed cell death processes often seen in signal transduction pathways. The pharmacodynamic parameters in the model are closely tied to the kinetics of tumor cell death, the potency of anticancer drugs, and the characteristics of tumor growth. These connections allow for an in-depth comparison of drug efficacy and provide insight into variations in tumor cell death processes.

This comprehensive model is articulated through the following system of differential equations:

\begin{equation}
\begin{aligned}\label{eq:pkpd}
    \frac{dx_1(t)}{dt} &= \frac{\lambda_0 \cdot x_1(t)}{[1 + \left(\frac{\lambda_0}{\lambda_1} \cdot w(t)\right)^{\Psi}]^{1/\Psi}} - k_2 \cdot c(t) \cdot x_1(t), \\
    \frac{dx_2(t)}{dt} &= k_2 \cdot c(t) \cdot x_1(t) - k_1 \cdot x_2(t), \\
    \frac{dx_3(t)}{dt} &= k_1 \cdot (x_2(t) - x_3(t)), \\
    \frac{dx_4(t)}{dt} &= k_1 \cdot (x_3(t) - x_4(t)), \\
    w(t) &= x_1(t) + x_2(t) + x_3(t) + x_4(t).
\end{aligned}
\end{equation}\\

In this model, $x_1(t)$ represents the fraction of proliferating cells, while $x_2(t)$ through $x_4(t)$ correspond to cells at different stages of damage leading to cell death. The total tumor mass, $w(t)$, accounts for all these cellular compartments, and $c(t)$ represents the plasma concentration of the anticancer drug. We note that while the drug's effect on tumor growth curves typically begins quickly, the observable impact often persists even after drug concentrations become negligible. This is because the rate-limiting step may lie in the kinetics of the transit compartment model. From the moment of inoculation (time zero) until  when drug exposure begins, the tumor experiences unperturbed growth. During this period, $c(t) = 0$ and $x_2(t) = x_3(t) = x_4(t) = 0$, meaning the total tumor mass $w(t)$ is equal to $x_1(t)$. \\

The drug concentration $c(t)$ is derived from a two-compartment model with predefined parameters, based on the data provided in \cite{simeoni2004predictive}, which was utilized in this study. The governing equations of the two-compartment PK model are described by the following system of linear ODEs:
\begin{equation}
\begin{aligned}
\label{eq:pk}
\frac{dA_1(t)}{dt} &= -k_{12} A_1(t) + k_{21} A_2(t) - k_{10} A_1(t) + I_1(t), \\
\frac{dA_2(t)}{dt} &= k_{12} A_1(t) - k_{21} A_2(t),\\
c(t) &= \frac{A_1(t)}{V_1}.
\end{aligned}
\end{equation}\\
 For the concentration of the drug in the PD model, we need to divide $A_1(t)$ by $V_1$, which is the volume of distribution in the first compartment. The PK parameters determined after the intravenous administration of paclitaxel are as follows: $V_1 = 0.81 , \text{L/kg}$, $k_{10} = 0.868 , \text{h}^{-1}$, $k_{12} = 0.0060 , \text{h}^{-1}$, and $k_{21} = 0.0838 , \text{h}^{-1}$. The administered dose is 30 mg/kg. The treatment starts on day 13, with a 3-dose regimen where 30 mg/kg is administered every 4 days.\\

\begin{figure}[!ht]
    \centering
    \includegraphics[width=0.8\linewidth]{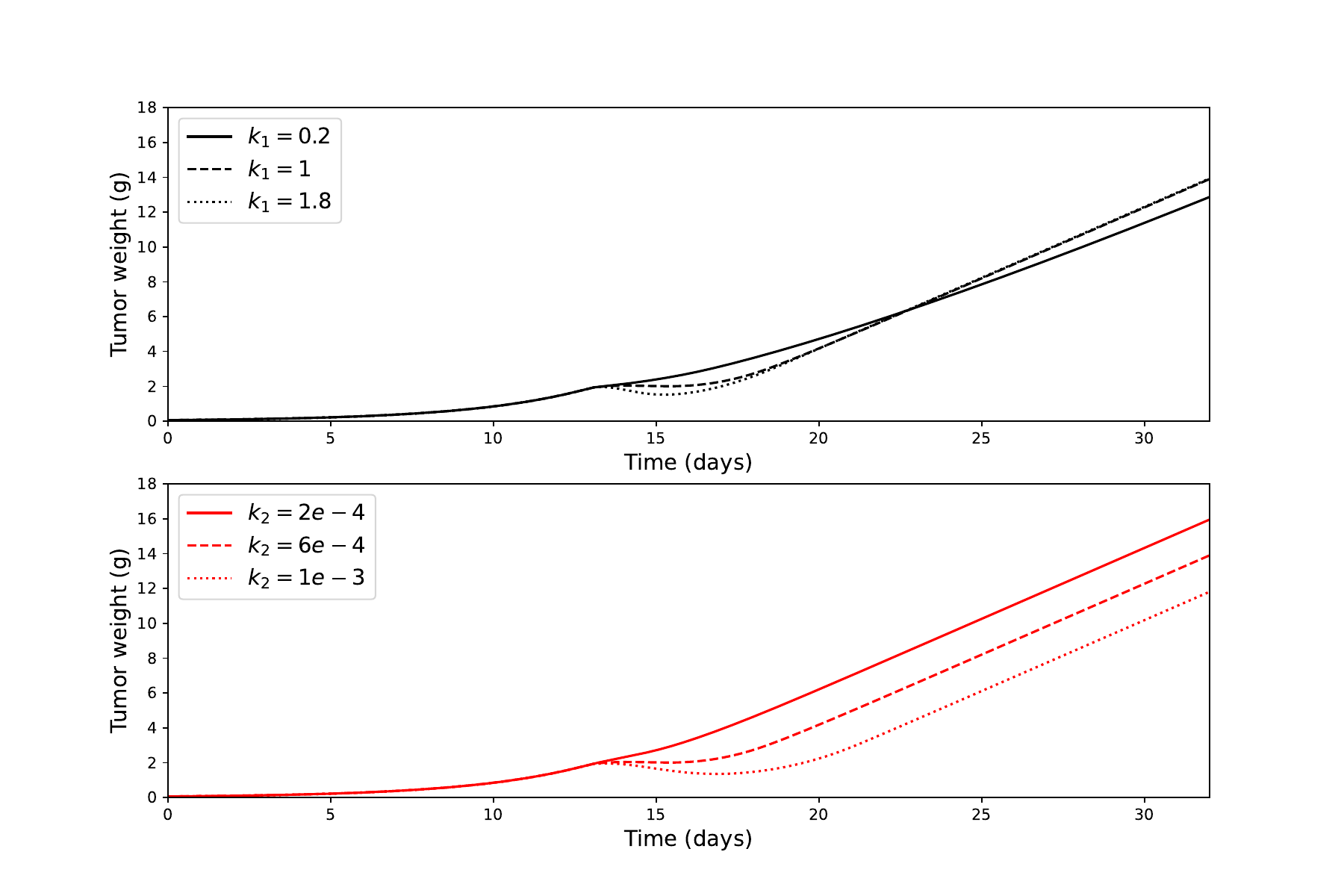}
    \caption{\textbf{Impact of \( k_1 \) and \( k_2 \) on tumor growth dynamics \eqref{eq:pkpd}.} In the top panel, \( k_2 \) is set to \( 6 \times 10^{-4} \) $\text{ml} \cdot {ng}^{-1}  \cdot \text{day}^{-1}$, while in the bottom panel, \( k_1 \) is set to 1 $day^{-1}$. The figure highlights the pharmacodynamic model's behavior following a single drug dose administered on day 13. Parameter \( k_1 \) represents the rate of drug-induced cell death. In contrast, \( k_2 \) quantifies drug potency, indicating how effectively the drug inhibits tumor growth. The figure reveals that increasing \( k_2 \) enhances treatment efficacy, resulting in a greater reduction in tumor size. In contrast, variations in \( k_1 \) yield only minor changes in the curve's overall shape and final slope.}

    \label{fig:k1_k2}
\end{figure}

Based on the CMINNs method, we generalized the system of four ODEs to a reduced compartment model with two ODEs by employing the compartment reduction block. This allows for finer control over the number of compartments while accurately modeling the delay by estimating the appropriate fractional order or inferring time-varying or piecewise constant value parameters. Further analysis is needed to determine which parameters should be treated as functions of time, constants, or piecewise constants. Figure \ref{fig:k1_k2} illustrates the effect-time profile of paclitaxel with varying parameter values in the proposed model. Parameter \( k_1 \) represents the rate of cell death, which varies over time due to the delay in cell death following drug administration. In contrast, \( k_2 \) serves as a measure of drug potency, reflecting the efficacy of the drug against the tumor. The figure shows that as the drug efficacy index \( k_2 \) increases, the treatment becomes more effective, leading to a further decrease in tumor weight. In contrast, changes in \( k_1 \) produce only slight variations in the curve's shape and the steady-state slope. Therefore, it is not necessary to model this parameter as time-dependent, as significant changes in \( k_1 \) have only a minor effect on the function's output.

Following the generalized modeling approach of CMINNs, we reduced the number of compartments in both physics-based models and applied the methods starting from the first drug administration. This decision is based on the observation that, in the absence of treatment, the model follows unperturbed tumor growth dynamics Eq.\eqref{eq:tumor}. Once treatment begins, these dynamics change. Therefore, it is logical to focus on persistence, resistance, tolerance, and delayed responses in cell death following the first drug exposure and solve the inverse problem for the system within the time range [13,32]. We modify the model as follows:

\begin{enumerate}

    \item In PINNs, due to the tolerance development of the multi-dose anti-cancer agent and the different behavior of cancer cells after repeated doses, we have designated \( k_2 \) as a piecewise constant parameter. This parameter will be recalculated from one administered dose to the next, allowing it to change after the second and third administrations. Meanwhile, \( k_1 \) is treated as a time-dependent function to better capture the dynamics of cell death. We reformulated the pharmacodynamic model as follows:
    \begin{equation}
    \label{eq:pd_pinns}
        \begin{aligned}
        \frac{dx_1(t)}{dt} &= \frac{\lambda_0 \cdot x_1(t)}{[1 + \left(\frac{\lambda_0}{\lambda_1} \cdot w(t)\right)^{\Psi}]^{1/\Psi}} \\
        &\quad -\sum_{i=1}^{2} k_{2,i} \cdot c_{(t_i, t_{i+1})}(t)\cdot x_{1(t_i, t_{i+1})}(t)  \\
        &\quad - k_{2,3} \cdot c_{(t_3, t_f)}(t) \cdot x_{1(t_3, t_f)}(t), \\
        \frac{dx_2(t)}{dt} &= \sum_{i=1}^{2} k_{2,i} \cdot c_{(t_i, t_{i+1})}(t)\cdot x_{1(t_i, t_{i+1})}(t) \\
        &\quad + k_{2,3} \cdot c_{(t_3, t_f)}(t) \cdot x_{1(t_3, t_f)}(t) \\
        &\quad - k_{1}(t) \cdot x_2(t), 
        \end{aligned}
    \end{equation}

    where $w(t) = x_1(t) + x_2(t)$ and $t_i$ denotes the $i$-th time of administration, with $i = 1, 2, 3$ for a three-dose administration regimen, and $t_f$ represents the final value in the time range over which the equation is solved. The parameter $k_{2,i}$ is a constant value of the piecewise constant parameter $k_2$ between the $i$-th and $(i+1)$-th administrations. The functions $c_{(t_i, t_{i+1})}(t)$ and $x_{1(t_i, t_{i+1})}(t)$ represent the parts of the function within the interval between $t_i$ and $t_{i+1}$.

    \item For fPINNs, to capture the memory effect and delayed response in cell death under a multi-dose treatment schedule, we fractionalized the PKPD equation \eqref{eq:pkpd} following our CMINNs method and the approach outlined by \cite{JK2022}. This method provides a more detailed description of the delay in drug response by incorporating fractional derivatives into the model. According to the paper, cellular response to drug administration is often delayed due to the age distribution of individual cells within a population. While transit compartment models (TCMs) are traditionally used to represent this delay, the proposed method introduces a single fractional derivative equation to model drug-induced damage, combining fractional and ordinary derivatives into one equation instead of using a system of ODEs across multiple compartments. A theorem that establishes rigorously such equivalence and corresponding theoretical foundations for this approach are detailed in \cite{JK2022}. The reduced fractionalized model is given by:

    \begin{equation}
    \begin{aligned}
    \label{eq:fpinnd}
    \frac{dx_1(t)}{dt} &= \frac{\lambda_0 \cdot x_1(t)}{[1 + \left(\frac{\lambda_0}{\lambda_1} \cdot w(t)\right)^{\Psi}]^{1/\Psi}} - k_2 \cdot c(t) \cdot x_1(t), \\
    \frac{dx_2(t)}{dt} &= k_2 \cdot c(t) \cdot x_1(t) - \tau^{-\alpha} D_t^{1-\alpha} x_2(t), 
    \end{aligned}
    \end{equation}
    where $w(t) = x_1(t) + x_2(t)$. The parameters $\lambda_0$, $\lambda_1$, and $\Psi$ represent tumor growth dynamics inferred from control data, while $k_2$, $\tau$, and $\alpha$ describe drug efficacy, the memory effect, and the delayed response in cell death following a multi-dose administration regimen. In this study, we set $\tau = 1$, consistent with the value of $k_1$ reported in \cite{simeoni2004predictive}. With this setting, we only need to infer the values of $k_2$ and $\alpha$ as model parameters using fPINNs.
\end{enumerate}

c\section{Results}\label{sec:results}
In this section, we present the results in tables that summarize the average inferred constant parameter values obtained after running the model five times with different random seeds for initialization of neural networks parameters. Alongside these averages, we report the standard deviations across the five runs. The corresponding ODE solutions, using the average inferred parameters, are illustrated in the figures. Although the neural network outputs provide the best-fit results—since minimizing data loss is its primary objective, standard numerical methods were employed to validate the results obtained from  solving the inverse problem by CMINNs method. This validation step is crucial in solving inverse problem by physics-based neural networks models because neural networks can achieve high data-fitting accuracy even when the inferred parameters are not entirely accurate. By solving the forward problem with numerical methods, we assessed the results using the inferred constant, piecewise constant, and time-varying parameters for the single-dose PK models (Models 1, 2, and 3) and the PD model (Model 4). For integer-order models, we used the \texttt{odeint} function from the \texttt{scipy.integrate} library, which utilizes the LSODA algorithm. LSODA dynamically switches between non-stiff and stiff solvers; for non-stiff problems, it applies the Adams-Bashforth-Moulton method (a multi-step approach), while for stiff problems, it employs the Backward Differentiation Formula (BDF), an implicit method. For simulating the PK model in Model 4, which involves a multi-dose drug administration regimen, we applied the explicit Euler forward method. For fPINNs, where we incorporate fractional-order derivatives, we used the FBDF method to compute the fractional derivatives within the explicit Euler forward method for solving the forward problem. The setup of model parameters for PINNs and fPINNs is reported in Table \ref{table:pinns_param} and Table \ref{table:fpinns_param}, respectively. Both methods demonstrate that \hyperref[sec:magin]{Model 2} and  \hyperref[sec:pkpd]{Model 4} present significant challenges for the PINNs and fPINNs approaches due to sparse data and abrupt changes in Model 2, and occasional spikes of drug administration in Model 4. To improve convergence and avoid local minima, we utilized a feature layer in both methods. For Model 2, we incorporated additional collocation points $N_{ode}$ in PINNs and added discretization points $N_{FBDF}$ for fractional derivatives. These enhancements increased the influence of the physics loss and allowed for a more accurate computation of the memory effect in areas with limited data, respectively.

\begin{table}[h!]
\centering
\begin{tabular}{|c|c|c|c|c|c|}
\hline
\textbf{Model} & \textbf{Arc.} & \textbf{\#Itr.} & \textbf{lr.} & \textbf{Feature Layer} & \textbf{$N_{ode}$} \\ \hline
Model 1 & 50,8 & 5000, 100000 & 1$e^{-3}$ & $\times$ & 140 \\ \hline
Model 2 & 20,4  & 10000, 100000 & 1$e^{-4}$ & $\checkmark$ & 650 \\ \hline
Model 3 (\ref{sec:3comp}) & 20,4 & 5000, 50000 & 1$e^{-3}$ & $\times$ & 300 \\ \hline
Model 3 (\ref{sec:3to2}) & 30,3 & 5000, 50000 & 1$e^{-3}$ & $\times$ & 300 \\ \hline
Model 4 & 30,6 & 30000/Interval & 1$e^{-4}$ & $\checkmark$ & 100/Interval \\ \hline
\end{tabular}
\caption{\textbf{PINNs parameter 
setup.} The first and second numbers in the 'Arc.' column refer to the width and depth of the neural networks, respectively. The initial and second numbers in the '\#Itr.' column represent the number of iterations during the primary and secondary training stages. 'lr' denotes the learning rate using the Adam optimizer. '$N_{ode}$' represents the number of collocation points. We used feature layers with lengths of 5 for models 2 and 4.}
\label{table:pinns_param}
\end{table}


\begin{table}[h!]
\centering
\begin{tabular}{|c|c|c|c|c|c|c|}
\hline
\textbf{Model} & \textbf{Arc.} & \textbf{\#Itr.} & \textbf{lr.} & \textbf{Feature Layer} & \textbf{$N_{FBDF}$} & \textbf{$N_{ode}$} \\ \hline
Model 1  & 20,4 & 300000 & 1$e^{-4}$ & $\times$&300 & 2000 \\ \hline
Model 2  & 20,3 & 300000 & 1$e^{-4}$ &$\checkmark$ &600 & 2000 \\ \hline
Model 3 (\ref{sec:3comp})  & 20,4 & 300000 & 1$e^{-4}$ &$\times$& 100 & 2000 \\ \hline
Model 3 (\ref{sec:3to2})  & 20,4 & 300000 & 1$e^{-4}$ &$\times$& 100 & 2000 \\ \hline
Model 4  & 30,6 & 200000 & 1$e^{-4}$ &$\checkmark$& 100 & 1900 \\ \hline
\end{tabular}
\caption{\textbf{fPINNs parameter setup.} The 'Arc.' column indicates the architecture of the neural networks, with the first and second numbers representing the width and depth, respectively. The '\#Itr.' column shows the number of training iterations. 'lr' refers to the learning rate used with the Adam optimizer. '$N_{FBDF}$' denotes the number of discretization points for the FBDF method, while '$N_{ode}$' specifies the number of collocation points. We used feature layers with lengths of 2 and 5 for models 2 and 4, respectively.}
\label{table:fpinns_param}
\end{table}


\subsection{Model 1}
In this model, we have only a single diffusion rate to explain the movement of mass to the second compartment and its subsequent return. The traditional method, with a constant value of $k$ in Eq. \ref{eq:gen12}, is unable to capture the data. The application of a single diffusion rate is inadequate for explaining the observed phenomena. Therefore, we utilize the CMINNs method to enhance the flexibility of the system and demonstrate how it transforms the compartmental modeling paradigm. This approach provides superior accuracy without the need for complex fractional mathematical models or the introduction of additional parameters to the model for more precise approximation of the experimental data.

\begin{figure}[!ht]
    \centering
    \includegraphics[width=0.6\linewidth]{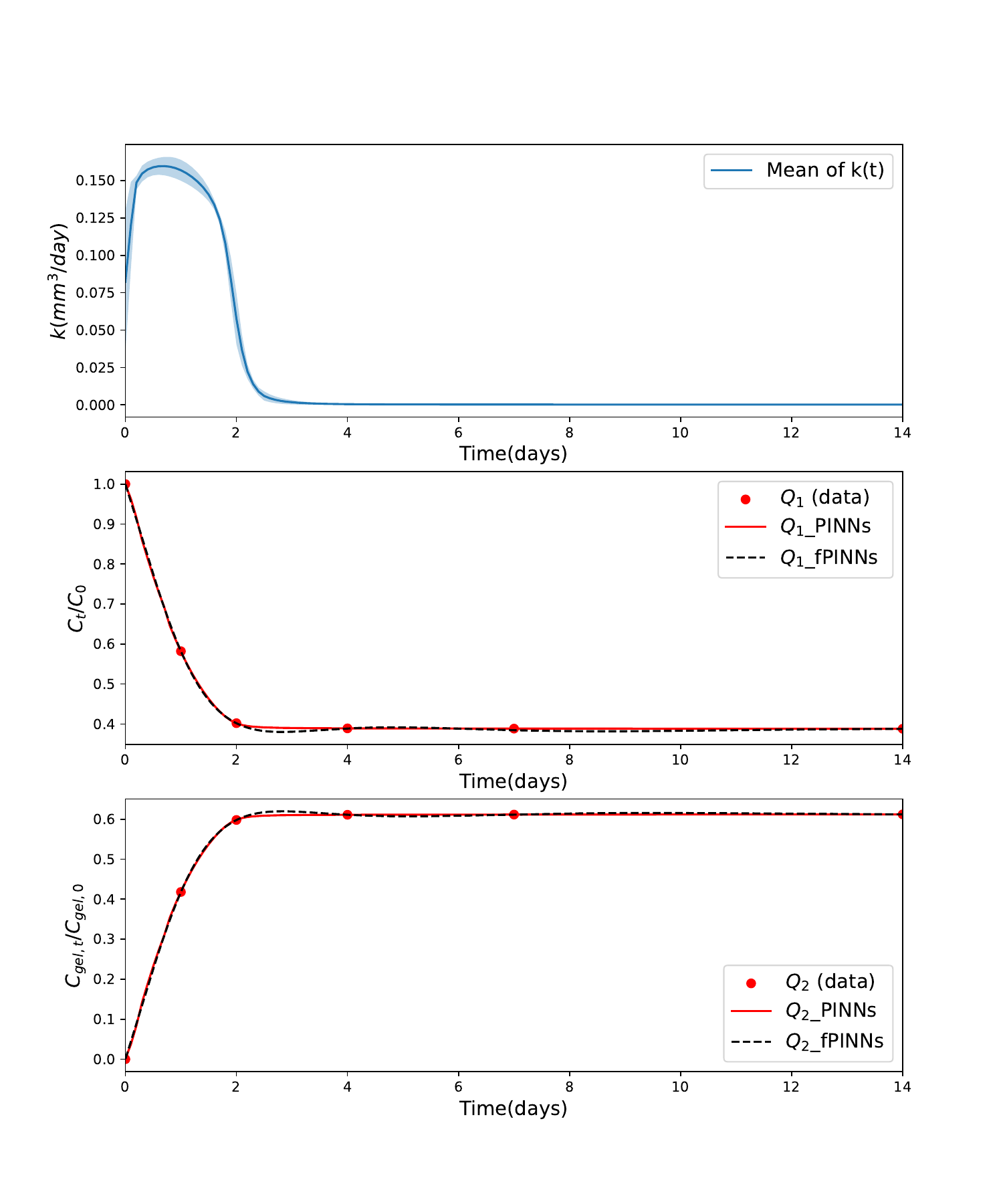}
    \caption{\textbf{Model 1. Comparison of the results between PINNs and fPINNs.} From top to bottom: the inferred time-varying parameter $k(t)$ in the PINNs method Eq. \eqref{eq:pinn_general}, the PINNs and fPINNs solutions for the first compartment, and the PINNs and fPINNs solutions for the second compartment. Total concentration of released gentamicin ($C_0$) and concentration of released gentamicin at time $t$ ($C_t$) are shown. $C_{\text{gel},t}$ represents the concentration of gentamicin remaining in the hydrogel at time $t$, while $C_{\text{gel},0}$ denotes the initial concentration of gentamicin inside the hydrogel. Data source: \cite{miskovic2023two}.}
    \label{fig:general}
\end{figure}

Miskovic et al. \cite{miskovic2023two} proposed a general fractional derivative of distributed order, which requires the addition of six parameters to the model. In contrast, we utilize the Caputo derivative to model the same phenomenon with the addition of a single parameter (fractional derivative order $\alpha$), and the degree of alignment between the results of these two approaches is demonstrated in Figure \ref{fig:general}. The final outputs of the PINNs method exhibit a smoother solution and offer new insights into the diffusion rate as a function. The model was run five times with different random seeds for the initialization of the neural network parameters, and the average of the inferred function, $k(t)$, was plotted. The light blue shading represents the standard deviation, while the blue line shows the mean value of $k(t)$). As the model does not include additional parameters influenced by the varying values of $k(t)$, the standard deviation is small, as reflected by the narrow light blue region. This indicates the robustness of our PINNs with time-varying parameter modeling. The mean and standard deviation of the fPINNs parameters are presented in Table \ref{tab:general}. In Figure \ref{fig:general}, the $k(t)$ function starts at approximately 0.050 $\text{mm}^3 / \text{day}$ at time $t=0$. As $k(t)$ increases, it reflects the acceleration of mass movement between the two compartments, with faster dynamics at the beginning. The subsequent decrease in $k(t)$ after 2 days indicates slower mass transfer. Around day 2.5, $k(t)$ stabilizes at a constant value of $k(t) = 0$. This is also observed in the concentration plots for the two compartments, which depict the same phenomenon: the mass in both compartments has reached equilibrium, and no further movement between compartments occurs.


\begin{table}[h!]
\centering
\begin{tabular}{c|c}
\hline
\textbf{Parameter} & \textbf{Mean $\pm$ Std} \\
\hline
$\alpha$ & 9.998 $e^{-1}$ $\pm$ 1.05$e^{-4}$ \\
$k$ & 9.54 $e^{-2}$ $\pm$ 2.26$e^{-3}$ \\
\hline
\end{tabular}
\caption{\textbf{Model 1: fPINNs 
results for identifying model 
parameters.} \(\alpha\) represents the fractional derivative order and \(k\) is the diffusion rate. The mean and standard deviation from 5 runs are presented.}

\label{tab:general}
\end{table}



\subsection{Model 2}
\begin{figure}[!ht]
    \centering
    \includegraphics[width=0.6\linewidth]{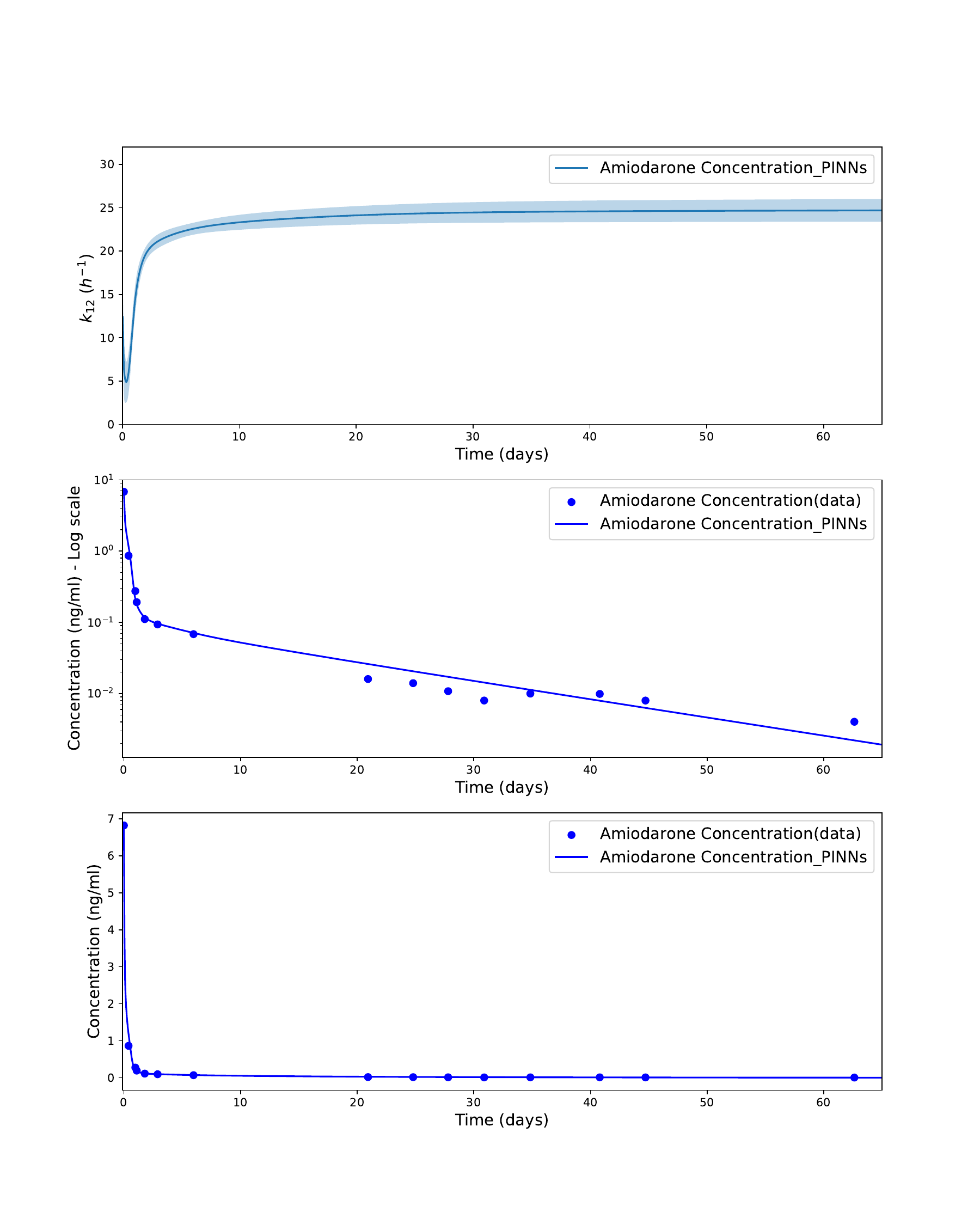}
    \caption{\textbf{Model 2. PINNs results.} The figure presents, from top to bottom: the inferred time-varying parameter \( k_{12}(t) \) using the PINNs method as described in Eq. \eqref{eq:pinns_magin}, the PINNs solutions on a logarithmic scale, and the PINNs solutions on a linear scale. This illustration highlights the effectiveness of PINNs in modeling drugs with anomalous decay behaviors, demonstrating a level of accuracy that traditional methods and integer-order derivatives cannot achieve. Data source: \cite{miskovic2023two}.}
    \label{fig:pinn_magin}
\end{figure}

In the CMINNs method using PINNs, the sparsity of data leads to multiple solutions with varying time-dependent functions for \( k_{12}(t) \), accompanied by slight variations in the coupled diffusion rate \( k_{21} \) values. This explains the larger standard deviations observed for the parameters \(k_{12}(t)\) in Figure \ref{fig:pinn_magin}  and \(k_{21}\) in Table \ref{table:results_magin}, both of which characterize the diffusion rate between compartments 1 and 2. Although each individual run produces slightly different results, incorporating these parameters into the numerical solver would yield a more accurate fit. Nonetheless, we opted to use the mean values of these parameters for solving the forward problem, while preserving the overall trend of the time-varying function rather than a specific function.

\begin{figure}[!ht]
    \centering
    \includegraphics[width=0.6\linewidth]{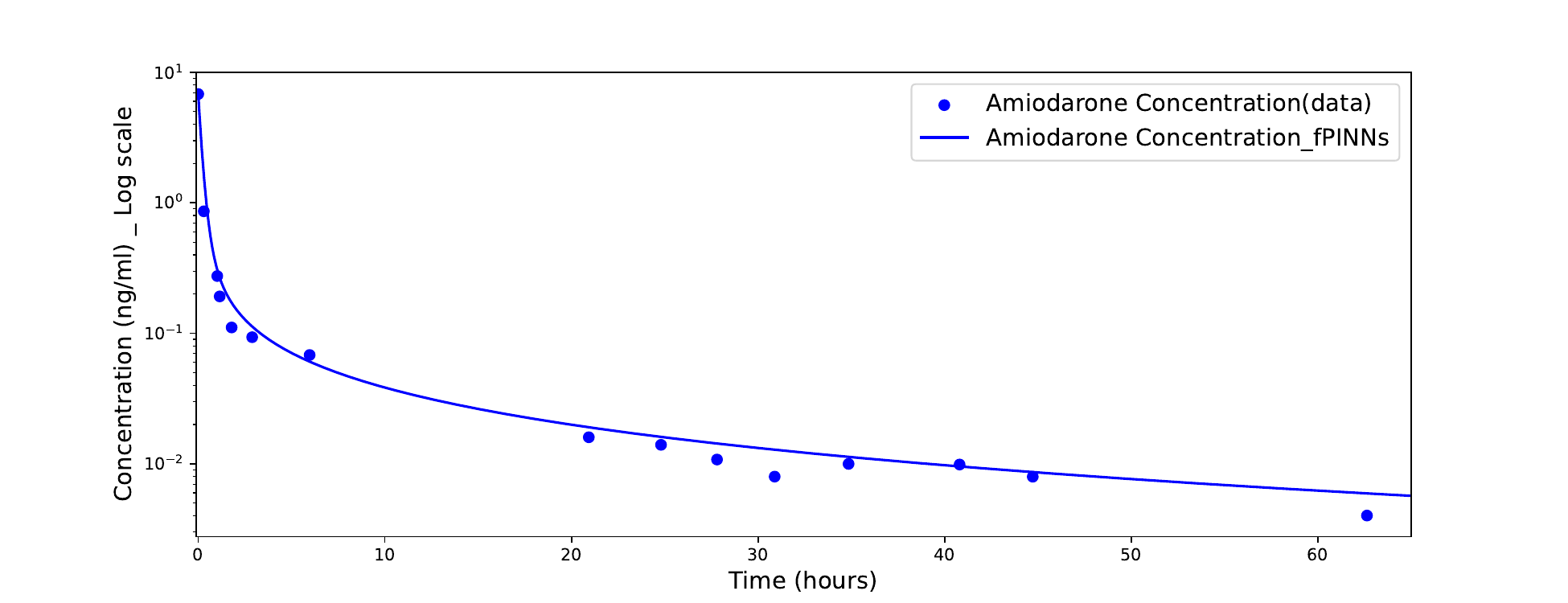}
    \caption{\textbf{Model 2: fPINNs results on logarithmic scale}. Fractional derivative models are commonly used in the literature to model the behavior of Amiodarone. The fPINNs method demonstrates excellent accuracy in optimizing the fractional order and model parameters.}
    \label{fig:fpinn_magin}
\end{figure}

Similarly, for the fPINNs, we employed the mean values as well, since this model represents the same phenomenon, and averaging was considered the most suitable approach to demonstrate robustness. The results are presented in Figure \ref{fig:pinn_magin} for the PINNs and Figure \ref{fig:fpinn_magin} for the fPINNs. The inferred parameter, denoted as \( k_{12}(t) \) , displays two distinct phases in its dynamics of absorption rate with the PINNs method. In the initial phase, the slope is markedly steep, indicating a rapid decline in the absorption rate during the initial two hours. Subsequently, as the drug accumulates in the plasma, the absorption rate increases suddenly, thereby accelerating the mass transfer between the central compartment and other organs in the peripheral compartment, leading to extensive tissue distribution. After a while, the drug concentration remains detectable in the plasma and continues to be either eliminated or redistributed to other compartments. Consequently, the absorption rate, represented by the parameter \( k_{12}(t) \), reaches a steady-state value. At this stage,\( k_{12}(t) \) is equal to a constant value, reflecting a consistent absorption rate for the remaining drug in plasma.

Comparing our results to those in the literature for Amiodarone, which exhibits non-exponential decay, the CMINNs approach demonstrated highly accurate results using an integer-order model. This outcome underscores the effectiveness of PINNs with time-varying parameters in modeling drugs characterized by anomalous decay behavior. Furthermore, Figure \ref{fig:pinn_magin} demonstrates the close alignment of the data points on a linear scale, which poses a significant challenge for many traditional methods relying on optimization techniques to solve inverse problems. However, on a logarithmic scale, small discrepancies in plasma drug concentration and model fitting errors become more apparent. This introduces an additional challenge for physics-based neural network methods, which must balance minimizing data loss with maintaining adherence to physical laws. These methods can sometimes converge to a local minimum, resulting in inaccurate inferred parameters, even though the neural network output fits the data perfectly. To address this, we introduced a feature layer as an auxiliary layer in the neural network to enhance the model. By incorporating parameters governing exponential decay, we improved the performance of both PINNs and fPINNs. The results from the fPINNs method demonstrate an even better fit, as it effectively captures the phenomenon of drug trapping and memory effect, providing a more comprehensive explanation for this issue and incorporating non-exponential decay phenomena.

\begin{table}[h!]
\centering
\begin{tabular}{c|c|c}
\hline
\textbf{Parameter} & \textbf{PINNs(Mean \(\pm\) Std Dev )}& \textbf{fPINNs(Mean \(\pm\) Std Dev )}\\ \hline
$\alpha$          & -                                 & 0.47  \(\pm\)4.09$e^{-3}$         \\
$k_{10}$          &1.70 \(\pm\) 6.39$e^{-7}$          & 1.77 \(\pm\) 2.16$e^{-2}$         \\
$k_{12}$          & Time variant                      &  4.30 \(\pm\) 4.45$e^{-2}$        \\
$k_{21}$          & 7.54$e^{-1}$\(\pm\)6.58$e^{-2}$  &  6.388$e^{-1}$ \(\pm\) 3.45$e^{-3}$        \\ \hline
\end{tabular}
\caption{\textbf{Model 2: PINNs and fPINNs results for identifying model parameters.} The elimination rate is nearly the same for both models. The time-varying parameter in PINNs does not significantly affect the variation of this parameter, but it does influence the standard deviation of the corresponding diffusion rate, \( k_{21} \).}

\label{table:results_magin}
\end{table}

\newpage
\subsection{Model 3}

For the Talaporfin sodium drug, we consider two different modeling approaches:
\begin{enumerate}[label=\roman*.]
    \item The first approach retains the three-compartment model but introduces greater flexibility to the existing model, allowing for a more accurate fit of the data across all compartments compared to traditional methods.
    \item The second approach utilizes the CMINNs method, which reduces the model to a two-compartment system and uses less data, considering only two compartments instead of three.
\end{enumerate}

\begin{figure}[!ht]
    \centering
    \includegraphics[width=0.6\linewidth]{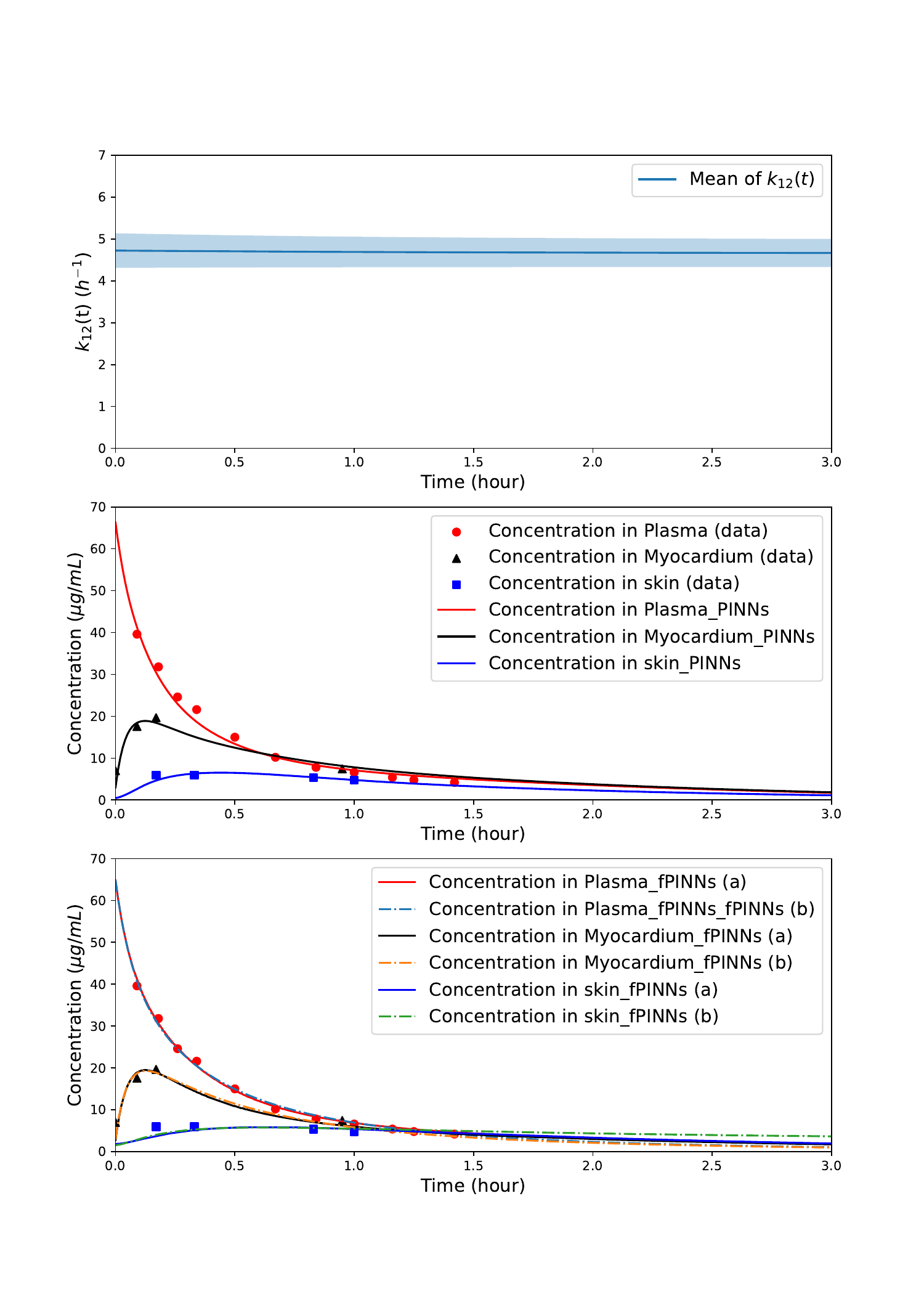}
    \caption{\textbf{Model 3. Comparison of PINNs and fPINNs results for the 
    refined three-compartment 
    models \ref{sec:3comp}.} From top to bottom: the inferred time-varying parameter $k_{12}(t)$ in the PINNs method, the PINNs solutions for the refined model, and the fPINNs solutions for the fractional models. Data source: \cite{bb0}.}

    \label{fig:3comp}
\end{figure}

The results for approach (i) are presented in Figure \ref{fig:3comp}, and the inferred parameter values are reported in Table \ref{tab:parameters_3comp}. We observe that, although the PINNs method allowed \(k_{12}(t)\) to vary with time, it exhibits only slight changes and consistently converges towards a nearly constant value. This observation validates our novel modeling approach: by reducing the number of compartments and subsequently the number of parameters, we maintain good model performance with enhanced control over fewer parameters. The constancy of \(k_{12}(t)\) can be attributed to the fact that the three-compartment model adequately captures the slope changes in the concentration-time profile. Furthermore, we observe a larger standard deviation in \(k_{21}\), indicating variability in the diffusion rate between compartments 2 and 1. When the rate of mass entering the second compartment increases, the rate at which it exits also increases. Despite these variations, the inferred parameter values are robust across all simulations and have comparably small standard deviation.We achieved a better fit compared to previous work \cite{bb0}, and we evaluated the fitting quality using the $R^2$ score. Our CMINNs approach demonstrates a fitting accuracy exceeding 0.99 for each individual run. This high accuracy is achieved without relying on the mean value typically used for plotting results, thereby confirming its superior performance. Additionally, we explore the impact of applying fractional derivatives to different sides of the model equations. In Table \ref{tab:parameters_3comp}, we observe that for compartments not involving fractional derivatives (i.e., the third compartment) and the elimination rate \(k_{10}\), the parameter values remain consistent across all models.\\


\begin{table}[ht]
    \centering
    \begin{tabular}{l|c|c|c}
        \hline
        \textbf{Parameter} & \textbf{PINNs (Mean \(\pm\) Std Dev)} & \textbf{fPINNs (a)} & \textbf{fPINNs (b)}  \\
        \hline
        \(\alpha\) & - & 9.917$e^{-1}$ & 6.74$e^{-1}$ \\
        \(k_{10}\) & 2.332 \(\pm\) 1.90$e^{-2}$  & 2.43 & 2.37  \\
        \(k_{12}\) & Time variant& 4.34 & 4.75  \\
        \(k_{21}\) & 9.27 \(\pm\) 5.85$e^{-1}$  & 10.02 & 10.85 \\
        \(k_{23}\) & 7.94 \(\pm\) 7$e^{-2}$ & 3.42 & 5.84  \\
        \(k_{32}\) & 4.11 \(\pm\) 2.60$e^{-2}$ & 1.33 & 3.44  \\
        \hline
    \end{tabular}
   \caption{\textbf{Model 3: Section \ref{sec:3comp}. Comparison of parameters for fPINNs (a), fPINNs (b), and PINNs.} fPINNs (a) and (b) represent commensurable and implicit non-commensurable fractional models, respectively. In (a), we employ a fractional derivative on the left side of the equation, while in (b), we fractionalize the right-hand side of the equation. In the PINNs method, we use a time-varying absorption rate \(k_{12}(t)\), and as expected, the corresponding diffusion rate shows the largest standard deviation compared to other parameters.}

    \label{tab:parameters_3comp}
\end{table}


\begin{figure}[!ht]
    \centering
    \includegraphics[width=0.6\linewidth]{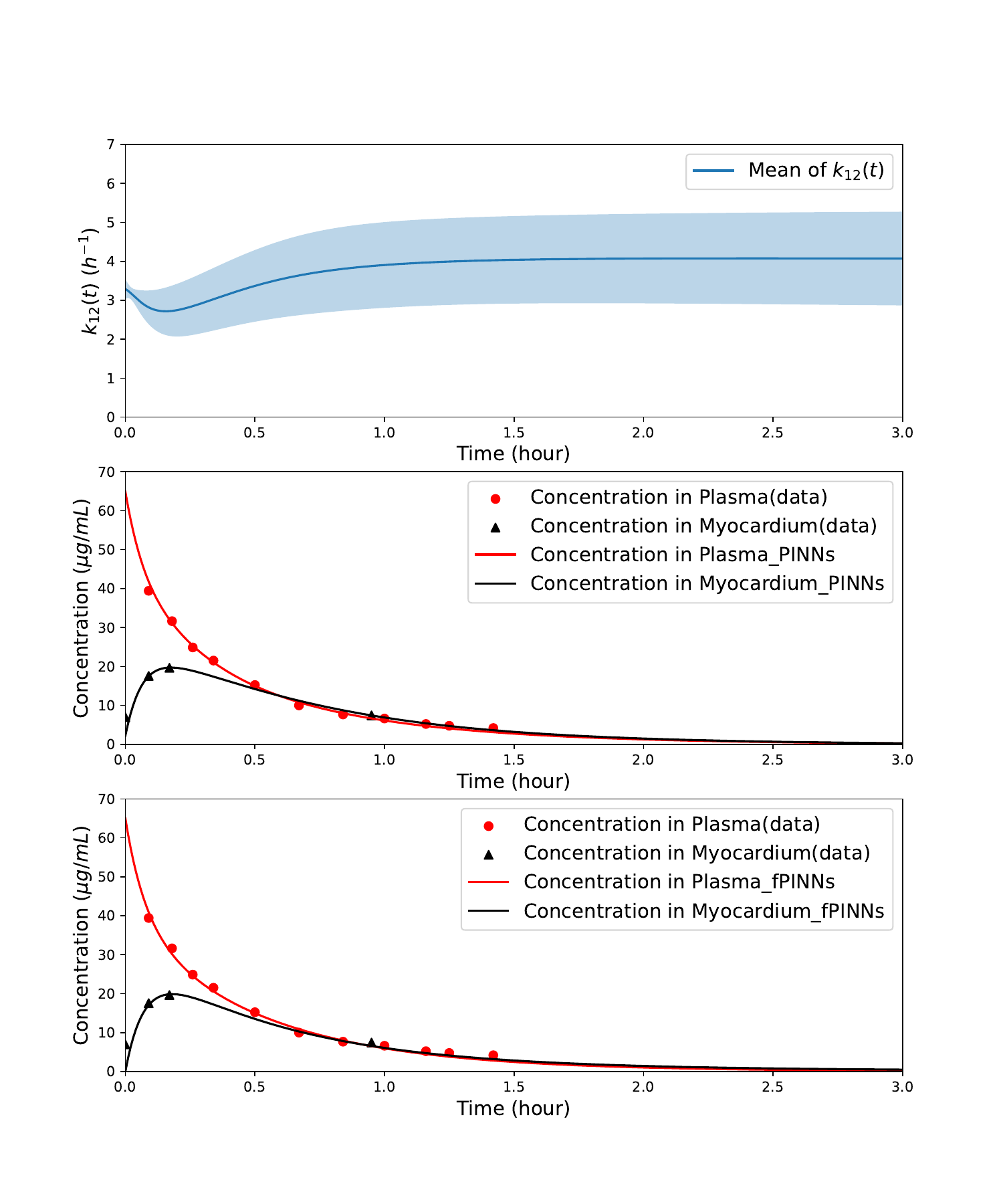}
    \caption{\textbf{Model 3. Comparison of PINNs and fPINNs results for the 
    reduced two-compartment 
    models \ref{sec:3to2}.} From top to bottom: the inferred time-varying parameter $k_{12}(t)$ in the PINNs method, the PINNs solutions for the reduced compartment model, and the fPINNs solutions for the reduced compartment fractional model.}
    \label{fig:3to2}
\end{figure}

The results for approach (ii), the reduced two-compartment model, are presented in Figure \ref{fig:3to2}. As a consequence of the reduction in the number of compartments, the value of $k_{12}(t)$ displays a recognisable trend over time. The process begins at 3.3, initially showing a decline, as substances diffuse from areas of higher to lower concentration, illustrating drug trapping in the peripheral compartment, which leads to a decrease in the absorption rate. As the drug redistributes to other organs, the absorption rate subsequently increases, eventually reaching a steady state that reflects a constant absorption rate. Our PINNs and fPINNs models fit the data accurately, even when using the mean inferred parameter values. The inferred parameter values are shown in Table \ref{table:3to2}. Since $k_{12}(t)$ shows significant variation across different runs, our elimination rate value, $k_{10}$, is also affected in this modeling. Although each individual set of inferred parameters from each run solves the problem, the differences in $k_{10}$ between the PINNs and 
fPINNs models highlight different solutions. Depending on the modeling approach we choose, the interpretation of the parameters can vary. Given that these models are not structurally identifiable, we can always obtain different solutions unless we constrain the search range for each parameter. Nonetheless, for $k_{12}(t)$, we consistently observe the same dynamics and trend, explaining the fluctuations in the absorption rate, regardless of its exact value.

\begin{table}[h!]
\centering
\begin{tabular}{c|c|c}
\hline
\textbf{Parameter} & \textbf{PINNs(Mean \(\pm\) Std Dev )}& \textbf{fPINNs(Mean \(\pm\) Std Dev )}\\ \hline
$\alpha$         & -  & 9.56$e^{-1}$\(\pm\) 2.07$e^{-2}$                  \\
$k_{10}$          & 3.42\(\pm\) 1.67$e^{-1}$  & 2.44 \(\pm\) 1.96$e^{-1}$                 \\
$k_{12}$          & Time variant & 3.27  \(\pm\) 4.10$e^{-1}$                 \\
$k_{21}$        & 7.64  \(\pm\) 6.35$e^{-1}$  & 8.895  \(\pm\) 1.194                \\ \hline
\end{tabular}
\caption{\textbf{Model 3: Section \ref{sec:3to2}. Comparison of Parameters for PINNs and fPINNs.} In this model, we reduced a three-compartment model to a two-compartment one and reformulated it for both PINNs and fPINNs. In the fPINNs method, we fractionalized the second compartment only on the left-hand side of the equation.}

\label{table:3to2}
\end{table}


\newpage
\subsection{Model 4}
Paclitaxel (PTX) is a widely used chemotherapy drug for treating various cancers. It targets microtubules, disrupting their dynamics to induce cell death. Despite its effectiveness, cancer resistance to PTX poses a significant challenge in clinical settings and is a leading cause of treatment failure and related mortality \cite{harris2006molecular, murray2012taxane, szakacs2006targeting}. Resistance to chemotherapeutic regimens has been reported for nearly all drugs used to treat the most lethal cancers. Unfortunately, drug resistance, tolerance, and persistence are often mistakenly regarded as synonymous, even though they are distinct phenomena, as shown in microbiology \cite{brauner2016distinguishing}. Persistence refers to the survival of a subpopulation during long-term drug exposure, even as the majority of the population is killed. Resistance is defined as the ability of an organism to grow at high drug concentrations, typically due to heritable mutations. Tolerance, on the other hand, refers to how the body handles the drug and how the drug distribution changes with repeated treatments. Tolerance is reversible, whereas resistance is irreversible.\\

\begin{figure}[!ht]
    \centering
    \includegraphics[width=0.65\linewidth]{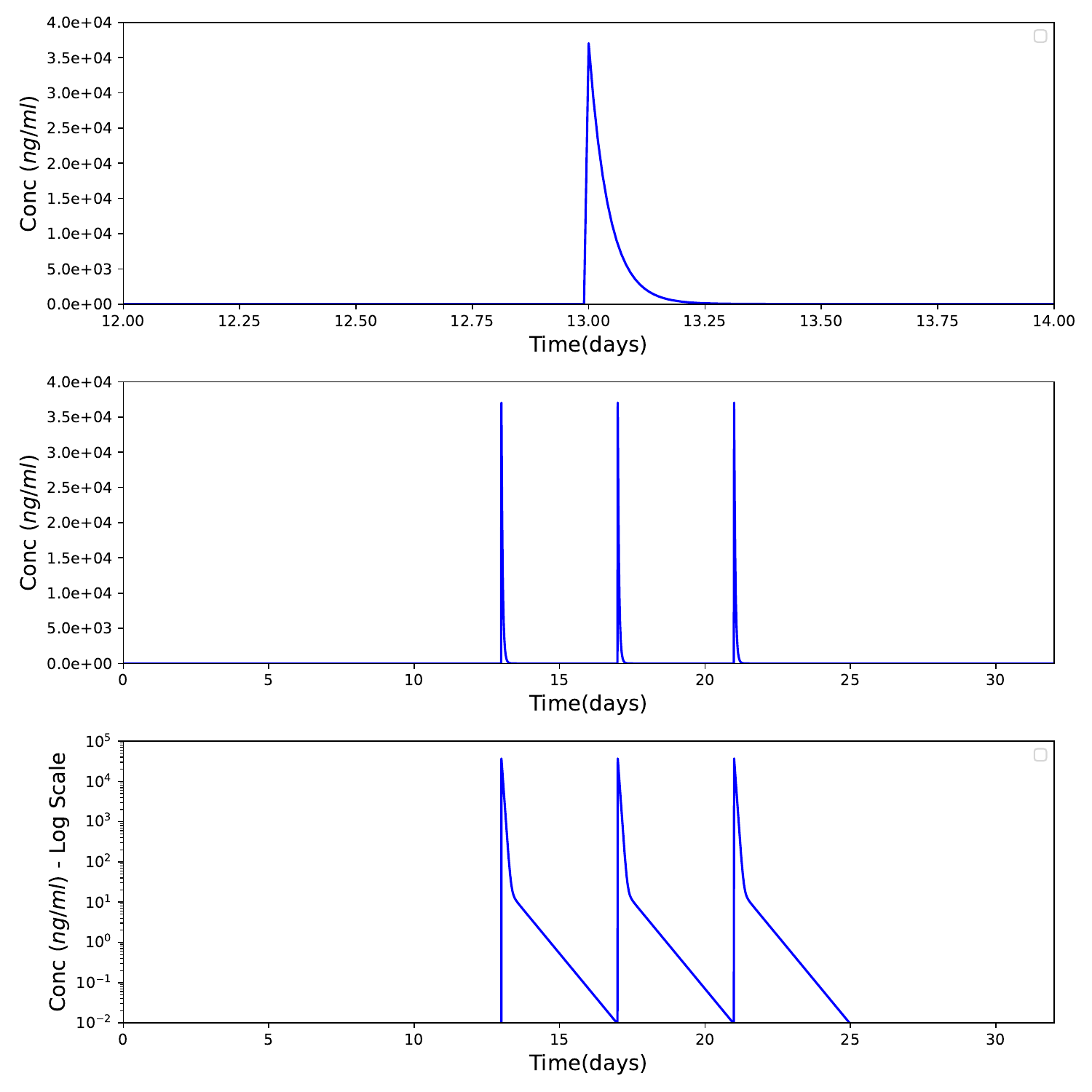}
    \caption{\textbf{Model 4. Comparison of paclitaxel concentration $c(t)$ in different scales from Eq. \eqref{eq:pk}.} In a three-dose treatment schedule, the drug is administered on days 13, 17, and 21. From top to bottom: single-dose $c(t)$ in linear scale, three-dose $c(t)$ in linear scale, and three-dose $c(t)$ in logarithmic scale.}
    \label{fig:pk}
\end{figure}

\begin{figure}[!ht]
    \centering
    \includegraphics[width=0.65\linewidth]{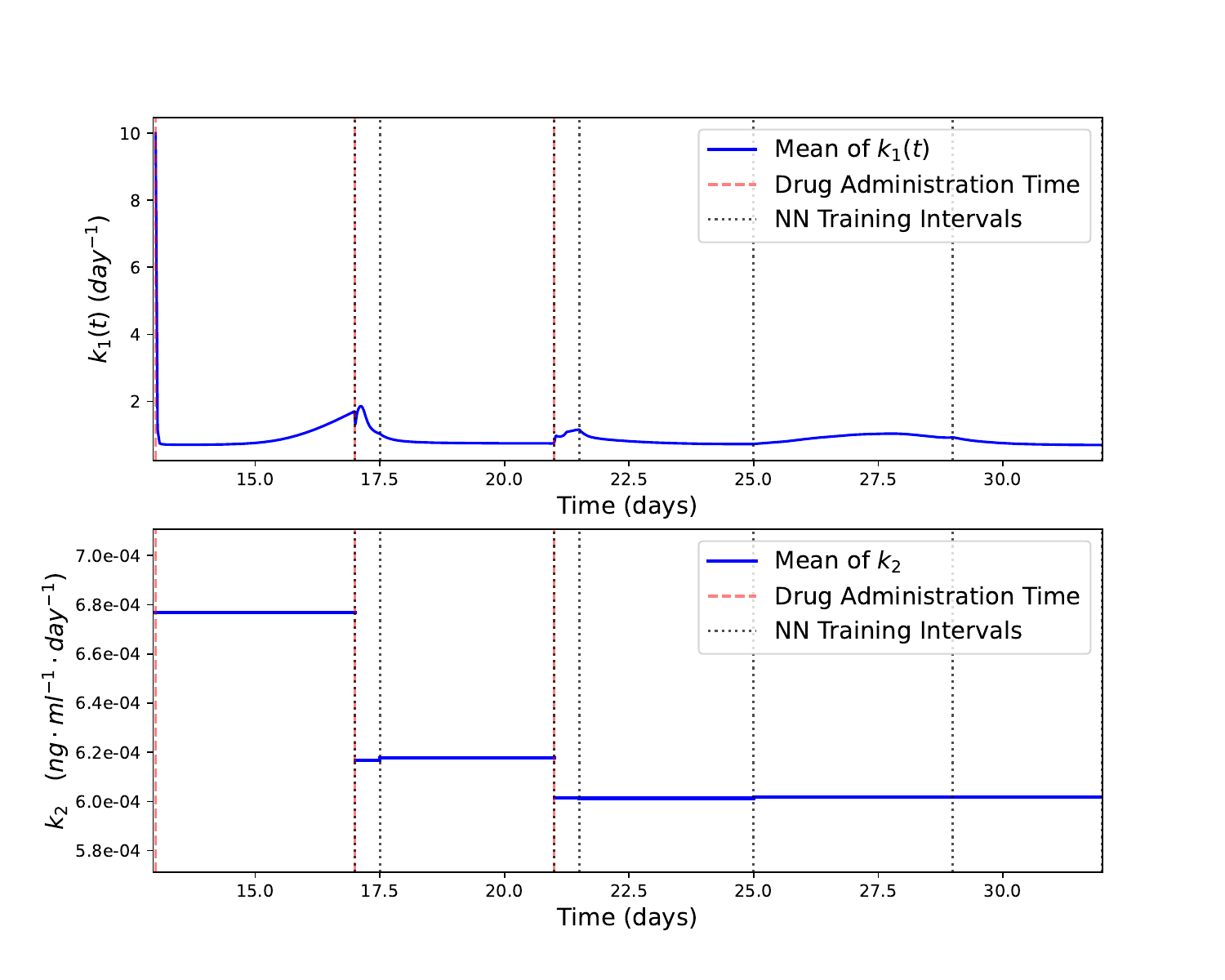}
    \caption{\textbf{Model 4. PINNs results for inferred $k_1(t)$ and $k_2$.} In a three-dose treatment schedule, the drug is administered every 4 days starting from day 13. From top to bottom: the inferred time-varying parameter $k_1(t)$ in Eq. \eqref{eq:pd_pinns}, and the inferred piecewise constant value of $k_2$.}
    \label{fig:k1k2}
\end{figure}

\begin{figure}[!ht]
    \centering
    \includegraphics[width=0.65\linewidth]{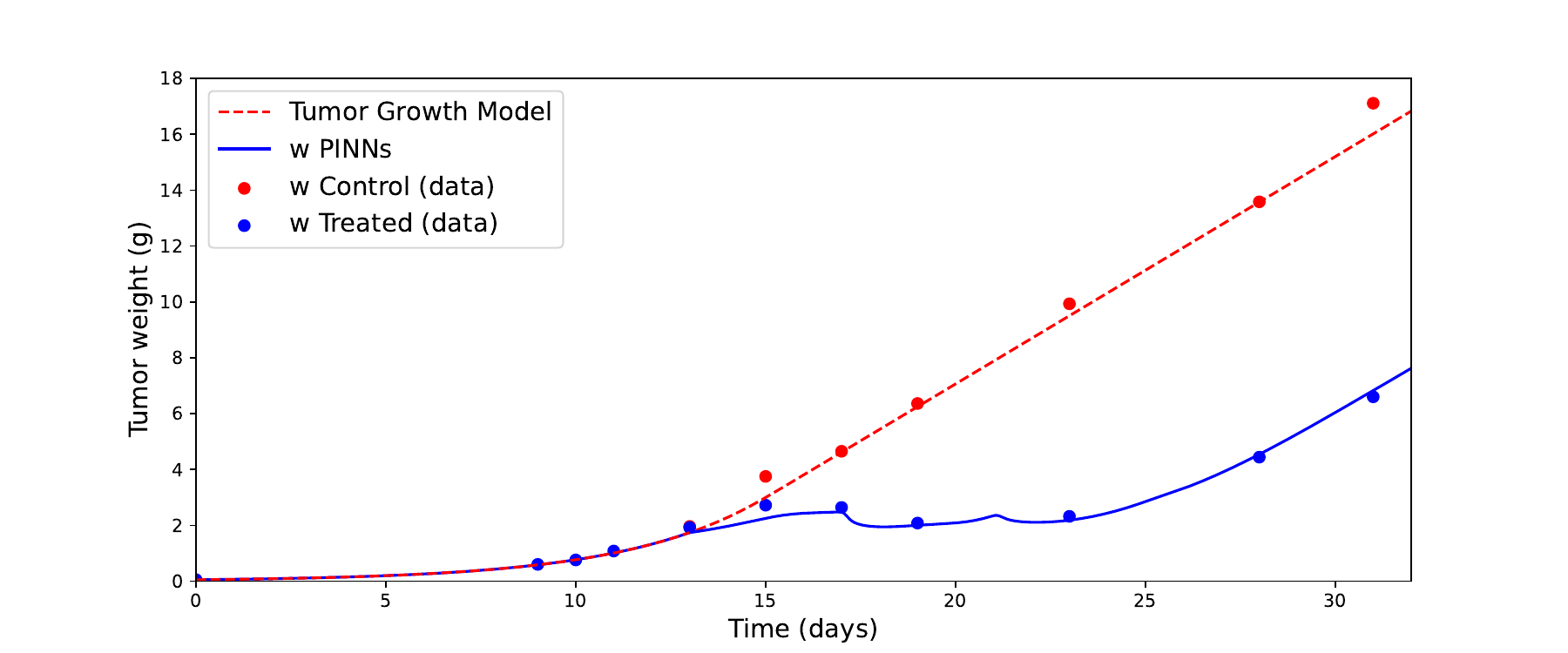}
    \caption{\textbf{Model 4. PINNs PD model solution with 
    inferred parameters $k_1(t)$ and $k_2$.} The red dashed line shows the tumor growth dynamic without treatment and blue curve is deviated from red dashed line on day 13, beginning of treatment. Data source: \cite{simeoni2004predictive}.}
    \label{fig:pd_pinn}
\end{figure}

\begin{figure}[!ht]
    \centering
    \includegraphics[width=0.65\linewidth]{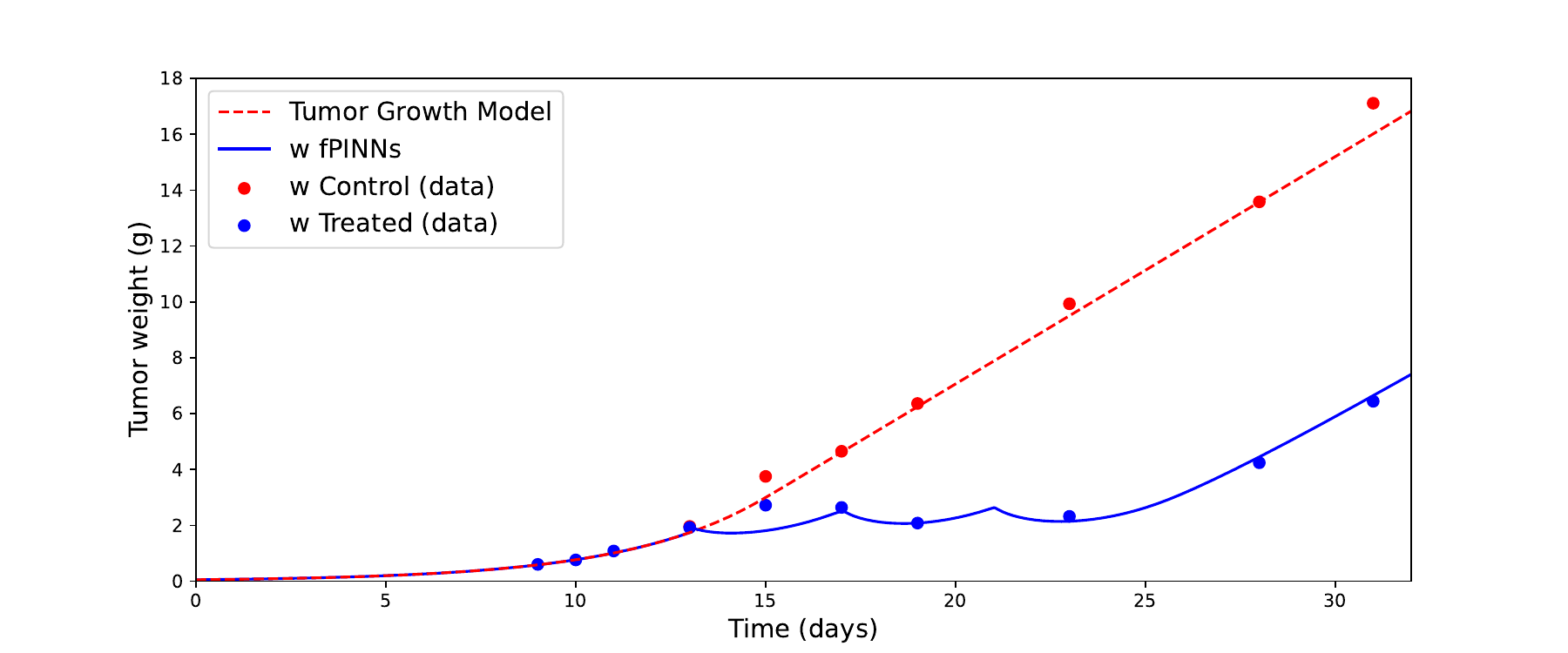}
    \caption{\textbf{Model 4. fPINNs PD model solution with inferred parameters $\alpha$ and $k_2$.}}
    \label{fig:pd_fpinn}
\end{figure}

The tumor growth model is analyzed under a three-dose chemotherapy regimen, as shown in Figure \ref{fig:pk}. We focus on the dynamics of the tumor growth variable $x_1(t)$, the number of dying cells $x_2(t)$, and the parameters $k_1(t)$ and $k_2(t)$, which describe the drug’s efficacy and the rate of cell death, respectively. For the tumor growth dynamic model, we used CMINNs to demonstrate how our novel modeling method with PINNs can capture various mechanisms from the reduced compartment model. By generalizing the compartment modeling approach, we can observe the dynamics of parameters in the model, which represents specific aspects of the phenomena. We trained our neural network over smaller intervals, beginning from the first dose of the drug. The intervals were set as \{[13, 16.99], [16.99, 17.5], [17.5, 20.99], [20.99, 21.5], [21.5, 25], [25, 29], [29, 32]\}. These intervals were selected to focus on capturing the spikes following drug administration, which occur around the same time as the treatment schedule (e.g., 13, 17, 21). The results for inferred parameters are shown in Figure \ref{fig:k1k2}, which displays the training intervals as well as the drug administration times. The rate at which damaged cells progress toward death, represented by \( k_1(t) \), demonstrated dynamic alterations in response to the multi-dose treatment regimen. Following the initial dose, \( k_1(t) \) exhibited a sharp spike, reflecting the rapid death of highly sensitive tumor cells. However, as treatment continued, the peak value of \( k_1(t) \) shows a reduction in magnitude and a slight increase in delay after each treatments. This gradual reduction and delay in \( k_1(t) \) over time indicate the emergence of resistance. While sensitive cells die off, the number of resistant cells increases, and these cells may generate even more resistant cells around the tumor, with some sensitive cells possibly hidden within it. The resistant tumor cell population requires a greater interval to accumulate lethal damage and die, resulting in smaller peaks between doses after the initial spike after each dose, which reflects the slower rate of damage in resistant cells. Furthermore, the increasing delay between doses highlights the adaptive response of the tumor to chemotherapy, with persistent cells continuing to grow and surviving even after treatment cessation. The parameter \( k_2 \), a piecewise constant index of drug efficacy, exhibits a decreasing trend across the treatment timeline, as shown in Figure \ref{fig:k1k2}. Although we did not restrict the training intervals to be only between treatments and included additional training intervals after each treatment until the next drug exposure, the value of \( k_2 \) is inferred to be constant unless a new drug administration occurred. After the first dose, \( k_2 \) was highest, indicating strong initial drug efficacy. However, subsequent doses result in progressively lower \( k_2 \) values, signaling a reduction in the drug's ability to inhibit tumor growth. This decline in \( k_2 \) suggests pharmacokinetic tolerance, where repeated drug administration leads to changes in the body's handling of the drug, such as enhanced metabolism or altered drug distribution, resulting in reduced drug concentrations at the tumor site. Consequently, the effective exposure of the tumor to the drug decreases, contributing to the observed reduction in \( k_2 \) and the diminished suppression of tumor growth. The decreasing \( k_2 \) values thus reflect both pharmacokinetic tolerance and the tumor's adaptive resistance to the chemotherapeutic agent. In Figure \ref{fig:pd_pinn}, the tumor weight, represented by $x_1(t)+x_2(t)$, exhibits a reduction following each dose of the chemotherapeutic agent, indicative of the drug's intended therapeutic effect. However, the magnitude of this reduction diminishes with each successive dose, particularly after the second and third administrations. This trend suggests the onset of pharmacodynamic tolerance or resistance, where tumor cells gradually become less responsive to the drug despite repeated exposure. The declining effect on tumor growth implies that adaptive mechanisms within the tumor may be at play, potentially involving the activation of survival pathways, mutations that confer drug resistance, or other cellular adaptations that decrease the drug’s efficacy over time. In Figure \ref{fig:pd_fpinn}, we present the results of tumor growth using the mean values of inferred parameters from fPINNs, as shown in Table \ref{table:pd_fpinss_stats}. \\
Both approaches show a good fit to the data, but PINNs provide more expressive and explainable results, offering new insights into the dynamics of adaptive mechanisms involved in chemotherapy.

\begin{table}[h!]
\centering
\begin{tabular}{c|c}
\hline
\textbf{Parameter} & \textbf{Mean \(\pm\) Std Dev} \\ \hline
$\alpha$ & 8.44$e^{-1}$ \(\pm\) 3.98$e^{-2}$ \\ 
$k_2$    & 6$e^{-4}$ \(\pm\) 3.46$e^{-7}$ \\  \hline
\end{tabular}
\caption{\textbf{Model 4. fPINNs 
inferred parameters.}}
\label{table:pd_fpinss_stats}
\end{table}

\newpage
\section{Summary and Discussion}\label{sec:conclusions}
The statistics of drug diffusion in the heterogeneous tissue does not follow normal distribution, hence the distributed delayed response cannot be faithfully  represented by the standard parameterized integer-order ODEs with constant parameter values. To better model this anomalous diffusion, herein we proposed two alternative approaches that can also reduce the effective number of ODEs. Rigorous theoretical work in the recent paper by \cite{JK2022} proved that a system of $(n+1)$ integer-order ODEs can be replaced by an equivalent system of $(1+1)$ fractional-order ODEs,  hence simplifying greatly the computational complexity of the problem while at the same time revealing the degree of anomalous diffusion via a single parameter, namely the order of the fractional time-derivative. Herein, we also provide an alternative method of reducing the number of compartments but introducing time-varying parameters in integer-order ODE systems. Unlike the fractional case, our proposal is empirical but it is justified by the results on different models we consider in our study. The implementation we propose based on PINNs and fPINNs is straightforward and can tackle even ill-posed problems.

Our work presents two significant contributions to pharmacokinetics and pharmacodynamics modeling. First, we applied the PINNs framework to a pharmacodynamic model with a multi-dose treatment schedule. To explore how the drug efficacy index changes after each dose, we solved the governing equations in smaller time intervals. This allowed us to examine variations in the constant parameter for drug efficacy, revealing that a decrease in this parameter corresponds to diminished drug effects on tumor growth. By employing the generalized two-ODE framework, our method demonstrated remarkable accuracy due to its flexibility in capturing multi-exponential and non-exponential decay phases. This led to explainable results that provide insights into the dynamics of absorption rates for different drugs, the memory effect that explains tissue trapping or delayed drug responses, and drug tolerance, resistance, and persistence in PD models for tumor growth. Our findings align with established literature, particularly regarding paclitaxel resistance.

Second, we introduced fPINNs to PKPD modeling, marking the first use of this approach in the field. By utilizing Physics-Informed Neural Networks, we automated the optimization of model parameters, especially fractional orders, as part of the training process. This innovation simplifies model adaptation, ensuring that the models achieve optimal alignment with empirical data across various methodologies. The use of fPINNs enhances the model's ability to capture complex dynamics, providing a more robust framework for understanding drug kinetics and responses. We tested various fractionalizing approaches and demonstrated that different models can yield equivalent results, though they offer different interpretations and aspects to consider in the modeling process.

In neural networks, catastrophic forgetting can occur, particularly in models addressing long time ranges, such as multi-dose administration in pharmacokinetic or pharmacodynamic models. This challenge can manifest when the network fails to capture one or more critical spikes early in the time series, often due to becoming trapped in local minima. For fPINNs, we addressed this issue by increasing the discretization of the FBDF, improving the model's capacity to capture these dynamics. Meanwhile, in PINNs, sequential learning—solving the ODEs in continuous intervals—resolves similar issues. The method proposed by \cite{kirkpatrick2017overcoming}, addresses the issue of catastrophic forgetting. Thus, exploring the incorporation of such techniques into CMINNs presents a promising direction for enhancing model performance across various datasets.

Although fPINNs are effective in handling sparse and noisy data sets, they have some drawbacks. First, convergence problems can arise due to optimization errors. In addition, using fPINNs as an inverse ODE solver is significantly more time consuming than using PINNs. However, when compared to conventional fractional derivative modeling methods, fPINNs provide an automatic, optimal approach. Therefore, future research should prioritize accelerating the training process of fPINNs for solving inverse ODE problems. Since the most time-consuming aspect is the computation of fractional derivatives by numerical methods, the development or use of faster numerical schemes within the fPINNs framework is a critical area for advancement.

Another challenge is managing small fluctuations in drug concentration values in plasma for pharmacokinetic model or tumor size within pharmacodynamic models over extended time periods. Values near zero and variations within this range can cause the outputs of fPINNs or PINNs to decay rapidly and approach zero, leading to inaccurate results for the inferred parameters. Additionally, sudden, large-scale changes in the data can shift the problem's scale too drastically, causing the neural network to become trapped in local minima and miss other important data points. This issue was addressed in this work by incorporating both a feature layer and a scaling layer into the network architecture. Furthermore, adopting a logarithmic scale for both the training dataset and the corresponding comaprtmental model could be a promising approach to better handle small variations and mitigate abrupt changes, thus enhancing the CMINNs performance across a broader range of scales. Another potential direction is integrating the CMINNs framework into state-space models like Mamba, which has been recently applied to predict tumor growth dynamics \cite{mamba}. By incorporating additional factors such as resistance, persistence, and tolerance dynamics in tumors—either as continuous functions or piecewise constants depending on repeated treatments—we may enhance the accuracy of predictions related to treatment efficacy and tumor behavior under different therapeutic strategies.
\section{Acknowledgements}
This work was supported by the National Institutes of Health (NIH) grant $R01HL154150$.
\bibliographystyle{ieeetr}
\bibliography{main}

\end{document}